\documentclass[aps,prb,reprint,floatfix,longbibliography]{revtex4-2}
\usepackage{graphicx}
\usepackage{amssymb}
\usepackage{amsmath,color}
\usepackage{bbold}
\usepackage{comment}
\pdfoutput=1
\usepackage[utf8]{inputenc}
\usepackage{physics}
\usepackage{amsfonts}
\usepackage{qcircuit}
\usepackage{tikz}
\usepackage{graphicx}
\graphicspath{ {./images/} }
\usepackage{floatrow}
\newfloatcommand{capbtabbox}{table}[][\FBwidth]
\usepackage{caption}
\usepackage{subcaption}
\usepackage{soul}

\begin{document}

\title{Simulated Non-Abelian Statistics of Majorana Zero Modes from a Kitaev Lattice}
\author{Foster Sabatino$^{1,2}$}
\author{Matthew Brooks$^3$}
\author{Charles Tahan$^{3}$}
\author{Silas Hoffman$^{1,4,5,6}$}
\affiliation{$^1$Department of Physics, University of Florida, Gainesville, FL 32611, USA}
\affiliation{$^2$Department of Physics, University of Central Florida, Orlando, Fl 32816, USA}
\affiliation{$^3$Department of Physics, University of Maryland, College Park, MD 20740, USA}
\affiliation{$^4$Quantum Theory Project,  University of Florida, Gainesville, FL 32611, USA}
\affiliation{$^5$Laboratory for Physical Sciences, 8050 Greenmead Drive, College Park, Maryland 20740, USA}
\affiliation{$^6$Condensed Matter Theory Center, Department of Physics, University of Maryland, College Park, MD 20742, USA}

\begin{abstract}
We simulate the non-Abelian exchange of Majorana zero modes (MZMs) on a quantum computer. Rather than utilizing MZMs at the boundaries of quantum Ising chains, which are typically represented as nonlocal operators on a quantum computer, using a Kitaev lattice allows us to exploit a local representation of MZMs. We detail the protocol for braiding two and four MZMs in terms of a spin Hamiltonian, i.e. physical qubit Hamiltonian. Projecting this onto a subspace of states, we extract an effective Hamiltonian which drives a non-Abelian Berry's phase. Using several approximations, we construct a set of gates which mimics this accumulation of non-Abelian phase and process this construction on a quantum computer. For two and four MZMs, we realize braiding fidelities of approximately 85\% and 47\%, respectively. 
\end{abstract}
 
\maketitle

\section{Introduction}

The many-body wavefunction of electrons acquires an overall minus sign upon exchanging any two constituent electrons. In contrast, anyons are so-named because an analogous many-body wavefunction can acquire a fixed root of negative one upon exchange \cite{nayakRMP08}. Moreover, some anyons, known as non-Abelian anyons, are equal to their original many-body wavefunction by a matrix multiplication.

One example of non-Abelian anyons are Majorana zero modes (MZMs). These particles were predicted to exist at the ends of one-dimensional topological superconductors \cite{kitaevPU01}. While the natural occurrence of such materials is unknown, several proposals have shown that such topological superconductors could be engineered using conventional materials \cite{lutchynPRL10,oregPRL10,aliceaPRB10}. While many experiments have shown signatures consistent with the presence of MZMs \cite{nadj-pergeSCI14,agheePRB23,agheeNAT25}, no definitive proof exists. Measurement of the non-Abelian statistics of MZMs would provide compelling evidence of their existence. This is, however, in general experimentally difficult.

In a completely different setup, MZMs are known to be emulated in quantum spin chains \cite{liebAoP61}. Taking advantage of the Jordan-Wigner mapping, one can show that MZMs are supported in spin-1/2 quantum Ising chains. However, owing to the nonlocal nature of the mapping, the MZMs are similarly nonlocal objects when mapped back onto the spin degrees of freedom. Nonetheless, some of the properties for which MZMs are well-known, e.g. perfect Andreev reflection \cite{hoffmanPRB23}, fractional Josephson junctions, and braiding, are present in these spin-emulated MZMs \cite{backensPRB17,stengerPRR21}. %Recent experiments have demonstrated braiding in this type of system.

In this work, we emulate Majorana fermions (MFs) inspired by the spin-1/2 Kitaev lattice \cite{kitaev2006anyons}. In contrast to the Ising chain, within this representation of MFs, the interactions between any pair of MFs can be emulated by local two-spin interactions. Consequently, to demonstrate braiding on a quantum computer emulating the spins, only two-qubit gates are necessary. In particular, we show that adiabatic manipulation of two-qubit interactions in a system of four qubits effectively braids two MZMs. Generalizing this system to ten qubits, we show a similar manipulation can braid four MZMs.

%This is most easily formulated 

%Demonstration of braiding is conveniently formulated using a `clock arm' Hamiltonian. 

%This allows us to construct a protocol for braiding MZMs using only two-qubit gates which can be used to show the inequivalence of braid order and, consequently, the non-Abelian nature of the MZMs. Additionally, one can `partially braid' two MZMs. 

%The four spins the Kitaev lattice, which can be equally well thought of as qubits, encode a single a logical qubit. With this encoding, one can effect partial braiding of the MZMs which correspond to arbitrary rotations about the $z$ axis of the Bloch sphere.  

%a similar protocol enables partial braiding of MZMs which correspond to arbitrary rotations around the $z$ axis of the Bloch sphere.

This paper is organized as follows: in Sec.~\ref{yjunc} we give the basic setup of the four and ten qubit systems, map it onto the equivalent MZM system, and calculate the non-Abelian gauge fields in both the reduced MZM space and spin space. In Sec.~\ref{sim} we map this protocol onto a set of quantum gates and run the protocal on a quantum computer. We discuss our results and provide an outlook in Sec.~\ref{disc} .

\section{Theoretical Model}
\label{yjunc}

%In this section, we calculate the Berry phases and the associated unitary operations that result from adiabatic manipulation of qubit coupling. It is convenient to formulate this in the language of logical qubits which are themselves composed of four physical qubits. 

%In this section, we review a model of four Majorana fermions (MFs), which can encode a single logical qubit. Exploiting a `clock arm' Hamiltonian, we show how adiabatic manipulation of their coupling can braid or partially braid two MZMs which results in an effective quantum gate acting on the logical qubits. We then consider a system of four qubits, i.e. two levels systems, and show that they can be mapped onto four coupled Majorana fermions. This can be generalized to ten MFs which can encode and couple two logical qubits using the same `clock arm' framework. %In both the one and two qubit cases, we explicitly show that adiabatic manipulation of the physical qubit coupling generates a Berry's phase on a logical qubit encoded in a two qubit subspace. 

%\subsection{Four Majoranas}
Consider a $Y$-junction of four Majorana fermions (MFs), $\gamma_j$ for $j=0,\ldots3$, where $\{\gamma_i,\gamma_j\}=2\delta_{ij}$ described by the Hamiltonian
\begin{align}
    \tilde H &= i[\Delta_z(\gamma_0 \gamma_1) + \Delta_y(\gamma_0\gamma_2)  + \Delta_x(\gamma_0\gamma_3) ] \nonumber \\
     &= i\gamma_0(\vec{\Delta} \cdot \vec{\gamma})\,.
     \label{Hmaj}
\end{align}

%When $\theta=0$, $\gamma_2$ and $\gamma_3$ commute with the Hamiltonian and are therefore MZMs. Upon parameterizing the coupling of the MFs by the coordinates on a sphere, $\vec\Delta=\Delta(\sin\theta\cos\phi,\sin\theta\sin\phi,\cos\theta)$ and fixing $|\vec\Delta|=\Delta$, one can generate 

\noindent These four MFs can be operated as a logical qubit by controlling the coupling of the MFs which we parameterize by the coordinates on a sphere, $\vec\Delta=\Delta(\sin\theta\cos\phi,\sin\theta\sin\phi,\cos\theta)$, fixing $|\vec\Delta|=\Delta$. When $\theta=0$, the system is idle and the MFs $\gamma_2$ and $\gamma_3$ commute with the Hamiltonian and are therefore MZMs. The Hamiltonian also commutes with the parity operator $\tilde n=i\gamma_2\gamma_3$ which has eigenvalues of $\pm1$, defining the logical state of the qubit. Moreover, there exists low- and high-energy subspaces according to the eigenvalue of $\tilde h=i\gamma_0\gamma_1=\mp1$, respectively; for fixed $\Delta$, Eq.~(\ref{Hmaj}) cannot mix low- and high-energy states. We choose to work in a subset of states that reside in the low energy subspace without loss of generality. One can perform a rotation about the $z$ axis of the Bloch sphere, using qubit states defined by the eigenvalues of $\tilde n$, by adiabatically changing $\theta$ and $\phi$ to trace a closed contour, $\Omega_c$, over the surface of a sphere. The resulting unitary operation is $\mathcal{R}_z(\Omega_c) = e^{-i\Omega_c\eta_z/2}$ with $\eta_z$ the Pauli matrix acting in the space of eigenvectors of $\tilde n$. When $\vec\Delta$ traces out an octant of the unit sphere with corners along the $x$, $y$, and $z$ axes, then $\Omega_c=\pi/2$ and one can show that this corresponds to braiding of $\gamma_2$ and $\gamma_3$. We refer to the case when $\Omega_c\neq\pi/2$ as `partially braiding' the MZMs.

\begin{figure}
    [b]
    \includegraphics[width=0.98\linewidth]{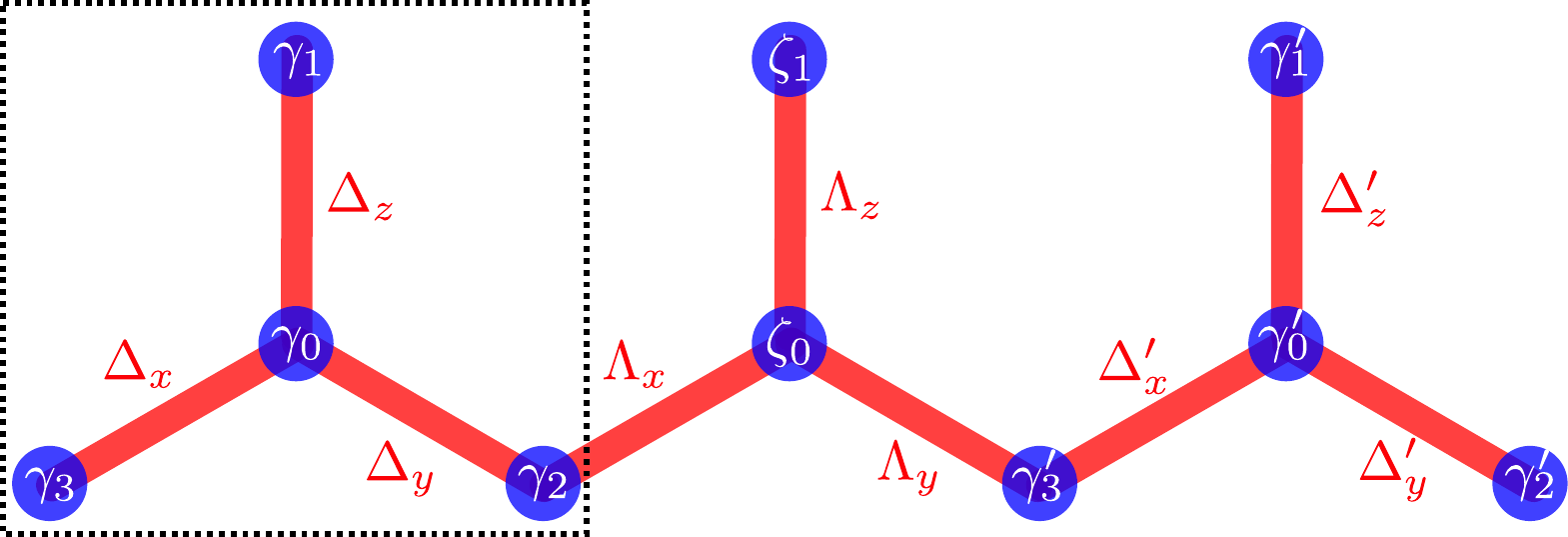}
    \caption{10 MZM, triple Y-junction system described in Eq.~(\ref{Hmaj10}). Each blue node houses the labeled MZM which are connected to the relevant neighbors by the red arms. Each arm is labeled by the relevant parameter of the clock arm vectors $\vec\Delta$, $\vec\Delta'$ and $\vec\Lambda$. The black dashed box highlights the Y-junction containing the first of the two topological qubits within this system and is equivalent to the system described in Eq.~(\ref{Hmaj})}
    \label{fig:MZM_diagram}
\end{figure}

Let us generalize the $Y$-junction to ten MFs wherein groups of three MFs are coupled through a central MF (Fig.~\ref{fig:MZM_diagram}),
\begin{equation}
    \tilde H_{10}=i\gamma_0(\vec\Delta\cdot\vec\gamma)+i\gamma'_0(\vec\Delta'\cdot\vec\gamma')+i\zeta_0(\vec\Lambda\cdot\vec\zeta)\,.
    \label{Hmaj10}
\end{equation}
Here, $\vec\gamma'=(\gamma_1',\gamma_2',\gamma_3')$ and $\vec\zeta=(\zeta_1,\gamma_3',\gamma_2)$ with $\gamma'_j$ and $\zeta'_j$ MFs obeying the usual $\{\zeta_i,\zeta_j\}=\delta_{ij}$, $\{\gamma'_i,\gamma'_j\}=\delta_{ij}$ and $\{\gamma_i,\gamma_j'\}=\{\gamma_i,\zeta_j\}=\{\gamma_i',\zeta_j'\}=0$. The parameters $\vec\Delta'=\Delta'(\sin\theta'\cos\phi',\sin\theta'\sin\phi',\cos\theta')$ and $\vec\Lambda=\Lambda(\sin\alpha\cos\beta,\sin\alpha\sin\beta,\cos\alpha)$ parameterize three clock arms. When $\theta=\theta'=\alpha=0$, the system is idle and $\gamma_2,\gamma_3,\gamma'_2$ and $\gamma_3'$ commute with the Hamiltonian and the energy is characterized by the eignevalues of $\bar h$, $\bar h'=i\gamma_2'\gamma_3'$, and $\bar h^a=i\zeta_0\zeta_1$. Analogous to the case of four MFs, one can perform unitary transformations on $n$ and $n'=i\gamma_2'\gamma_3'$ by moving the clock arms. It is easy to see that adiabatically moving $\vec\Delta'$ over the unit sphere, tracing a closed contour of area $\Omega_c$ generates the operation $\mathcal R_z'(\Omega_c)=\exp(-i\Omega_c\eta'_z/2)$ where $\eta'^\alpha$ are the Pauli matrices acting on the space of states with $i\gamma_2'\gamma_3'=\pm1$. Adiabatically tracing a contour of area $\Omega_c$ on the unit sphere with the vector $\vec\Lambda$ performs the operation $\mathcal R_{xx}(\Omega_c)=\exp(-i
\Omega_c\eta_x\eta_x'/2)$. By moving $\vec\Delta'$ ($\vec\Lambda$) in a loop enclosing an area of $\Omega_c=\pi/2 $on the unit sphere, $\gamma_2'$ and $\gamma_3'$ ($\gamma_2$ and $\gamma_3'$) are braided.

%Consider the states $|s,s'\rangle$ to be eigenstates of $i\gamma_2\gamma_3$ and $i\gamma_2'\gamma_3'$ with eigenvalues $s$ and $s'$, respectively. Let a us choose the basis states of a logical qubit to be $|-1,-1\rangle$ and $|1,1\rangle$. Notice that partially braiding $\gamma_2$ and $\gamma_3$ or $\gamma_2'$ and $\gamma_3'$ generates a rotation about the $z$ axis of the Bloch sphere. Paritally braiding $\gamma_2$ and $\gamma3'$ generates a rotation about the $x$ axis of the Bloch sphere. Consequently, adiabatic control of the coupling of MFs grants access to any point on the Bloch sphere.

In order to emulate the MFs in Eq.~(\ref{Hmaj}), consider three qubits whose coupling to a central qubit are noncommuting,
\begin{equation}
     H = \Delta_z\sigma_0^z\sigma_1^z + \Delta_y\sigma_0^y\sigma_2^y + \Delta_x\sigma_0^x\sigma_3^x \,.
     \label{ham4}
\end{equation}
We recognize this as a piece of the so-called Kitaev lattice which itself is known to support dispersing Majorana edge modes. Inspired by the latter, we extend the dimension of the space by writing the Pauli matrices as products of Majoranas, $\sigma^{\alpha}_{i} = i\gamma^{\alpha}_{i}\gamma_{i}$, where $\alpha=x,y,z$, $i=1,2,3,4$. Here, $\{\gamma^\alpha_i,\gamma^\beta_j\}=\delta_{ij}\delta_{\alpha\beta}$, $\{\gamma_i,\gamma_j\}=\delta_{ij}$, and $\{\gamma_i^\alpha,\gamma_j\}=0$. Upon appropriate substitution (see appendix), one can show that $H$, $h=\sigma_0^z\sigma_1^z$, $n=\sigma_2^z\sigma_3^z$ can be mapped into $\tilde H$, $\tilde h$, and $\tilde n$, respectively. Moreover, there exist two equations of motion, $W_1=\sigma^z_0\sigma^x_2\sigma^y_3$ and $W_2=\sigma^y_0\sigma^x_1\sigma^z_3$, which commute with $H$ and $n$; we henceforth restrict to the subspace of states which are eigenvectors of $W_1$ and $W_2$ with eigenvalue $-1$. Within this subspace, Eq.~(\ref{ham4}) becomes (see Appendix)
\begin{equation}
    H_\textrm{eff}=-\tau^z\cos\theta-\tau^x\eta^x\sin\theta\cos\phi+\tau^x\eta^y\sin\theta\sin\phi\,,
    \label{Heff}
\end{equation}
where $\tau^\alpha$ for $\alpha=x,y,z$ are the Pauli matrices acting in the space of low and high energy. We find that the gauge potentials associated with adiabatic change of $\theta$ and $\phi$ are
\begin{align}
    \mathcal A_\theta&=-i\tau^y\eta^y/2\,,\nonumber\\
    \mathcal A_\phi&=i(\eta^z\cos\theta-\tau^y\eta^x\sin\theta)/2\,,
\end{align}
respectively (see Appendix). A unitary rotation, $U=\mathcal P \exp(-\int \mathcal A_\theta d\theta+ \mathcal A_\phi d\phi)$, is generated upon moving around a closed loop on the unit sphere. Although the path-ordering makes this difficult to calculate in general, when the closed moves for the $z$ axis, to the $x$ axis, around the equator by and angle $\varphi$ and back to the $z$ axis, the area enclosed by the loop is $\Omega_c=\varphi$ and $U=\exp(-i\varphi\eta^z/2)$ which matches the partial braiding of MFs.  

%\subsection{Ten Majoranas}
Analogous to four Majoranas, Eq.~(\ref{Hmaj10}) can be emulated with ten physical qubits according to
\begin{align}
    H_{10}&= \Delta_z\sigma_0^z\sigma_1^z + \Delta_y\sigma_0^y\sigma_2^y + \Delta_x\sigma_0^x\sigma_3^x\nonumber\\
    &+\Delta_z'\sigma_{0'}^z\sigma_{1'}^z + \Delta_y'\sigma_{0'}^y\sigma_{2'}^y + \Delta_x'\sigma_{0'}^x\sigma_{3'}^x\nonumber\\
    &+\Lambda_z\sigma_4^z\sigma_5^z +\Lambda_y\sigma_4^y\sigma_{3'}^y+\Lambda_x\sigma_2^x\sigma_4^x\,.
    \label{H10m}
\end{align}
Upon rewriting the Pauli matrices as products of Majoranas and using constants of motion, analogous to the procedure for four physical qubits, one can show that Eq.~(\ref{H10m}) maps onto Eq.~(\ref{Hmaj10}) (App.~\ref{app:Kitaev_lattice}). Similarly, one can show that $n'=\sigma_{2'}\sigma_{3'}$, $h'=\sigma_{0'}^z\sigma_{1'}^z$, $h^a=\sigma_4^z\sigma_5^z$ map onto $\tilde n'$, $\tilde h'$ and $\tilde h^a$, respectively. In addition to $W_1$ and $W_2$, there are three additional constants of motion, $W_4=\sigma^z_{0'}\sigma^x_{2'}\sigma^y_{3'}$, $W_5=\sigma^x_{0'}\sigma^y_{1'}\sigma^z_{2'}$, and $W_6=\sigma^x_0\sigma^y_1\sigma^z_2\sigma_4^y\sigma_5^y$ which commute with $H_{10}$. Without loss of generality, we restrict to the space of states in which the constants of motion are $-1$.

For our purposes, it is convenient to rotate only one clock arm Eq.~(\ref{H10m}) while the other two remain idle. When the middle and right clock arms are idle, i.e. $\theta'=\alpha=0$, we recover Eq.~(\ref{Heff}). When the left and middle clock arms are idle, i.e. $\theta=\alpha=0$, we obtain the effective Hamiltonian (see Appendix)
\begin{equation}
    H_\textrm{eff}'=-\tau^{z\prime} \cos\theta'-\tau^{x\prime} \eta^{x\prime}\sin\theta'\cos\phi'+\tau^{x\prime} \eta^{y\prime} \sin\theta'\sin\phi'\, ,
\label{Heff'}
\end{equation}

\noindent where $\tau^{\alpha\prime}$ acts on the high- and low- energy subspace corresponding to the eigenvalues of $h'=\pm1$, respectively. Fixing the left and right clock arms in the idle state, i.e. $\theta=\theta'=0$, we generate the effective Hamiltonian (see Appendix),

\begin{equation}
    \begin{split}
        H_\textrm{eff}^a= -\chi^z\cos\alpha+\chi^x\eta^y\sin\alpha\cos\beta&\\
            +\chi^y\eta^{x\prime}\sin\alpha\sin\beta&\,,
    \end{split}
\end{equation}

\noindent where $\chi^\alpha$ acts on the space of states corresponding to the eigenvalues of $h^a$. It is straightforward to show that the gauge potentials generated by changing the right clock arm are 
\begin{align}
    \mathcal A_{\theta'}&=-i\tau^{y\prime}\eta^{y\prime}/2\,,\nonumber\\
    \mathcal A_{\phi'}&=i(\eta^{z\prime}\cos\theta'-\tau^{y\prime}\eta^{x\prime}\sin\theta')/2\,,
\end{align}

\noindent and that one can generate a unitary rotation $U=\exp(-i\varphi\eta^{z\prime}/2)$ when the clock arm, starting from idle, follows a path to the equator, around the equator by $\varphi$ and back to idle. Similarly, one can show that the gauge potentials generated by changing the middle clock arm are
\begin{align}
    \mathcal A_{\alpha}&=i\chi^x\eta^x/2\,,\nonumber\\
    \mathcal A_{\beta}&=i(\chi^z\eta^x\eta^{x\prime}\cos\alpha - \chi^y\eta^{x\prime}\sin\alpha)/2\,.
\end{align}
Following the same path as for the previous clock arm, one generates the unitary rotation $U=\exp(-i\varphi\eta^x\eta^{x\prime}/2)$.

\section{Simulated Braiding}
\label{sim}

\begin{figure}
    [b]
    \includegraphics[width=0.49\linewidth]{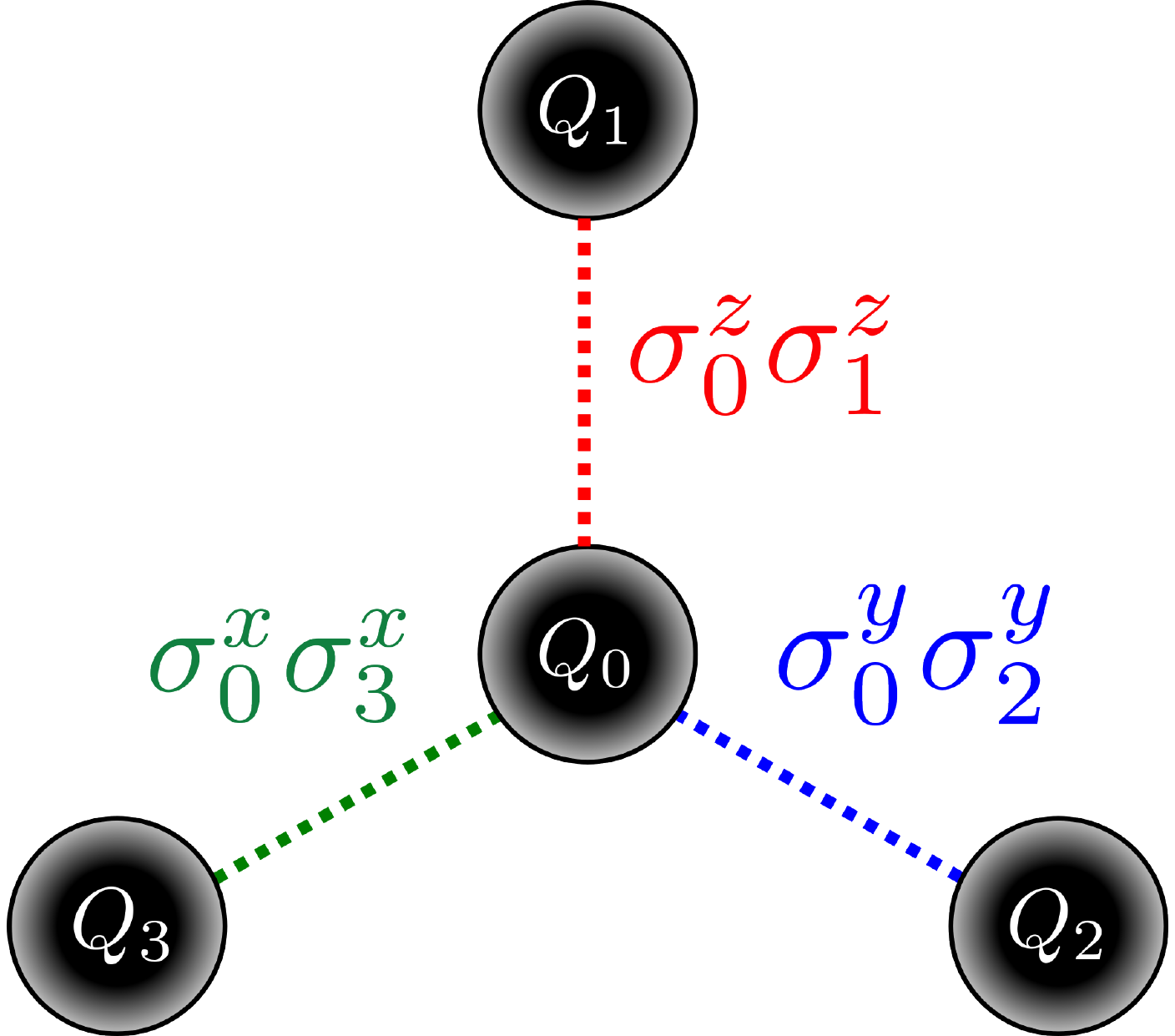}
    \includegraphics[width=0.49\linewidth]{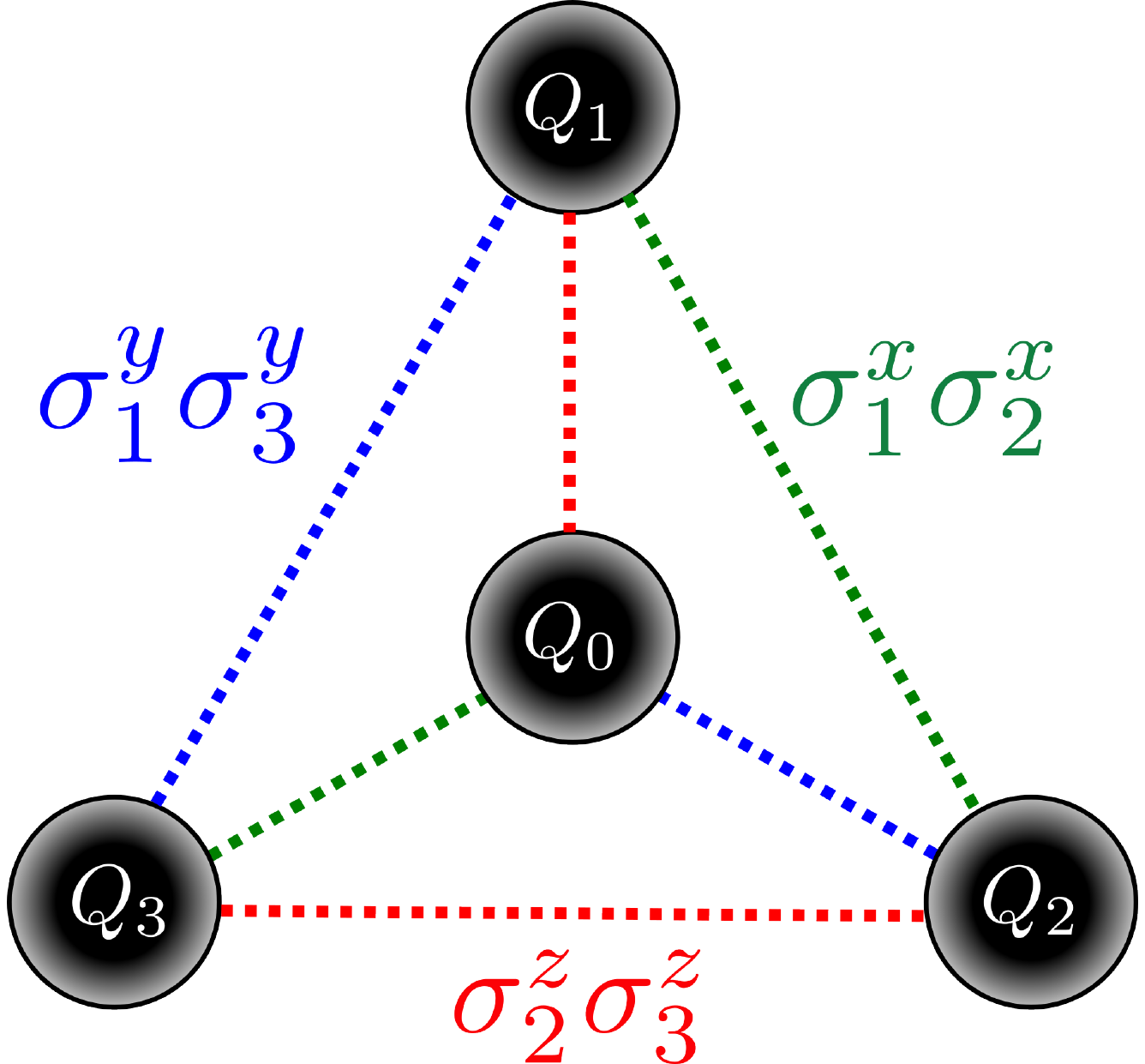}
    \caption{Qubit analogue of a four MZM Y-junction with Kitaev lattice like connectivity. 
    %(b) Qubit analogue of  2 4-MZM Y-junction qubits linked  by a third coupling Y-junction introduced by the addition of two ancilla qubits.
    }
    \label{fig:SingleQubit}
\end{figure}

We proceed to simulate braiding of MZMs on a quantum computer. In this section, we detail the initialization of the logical state and the necessary gate operations to evolve the states according to Eqs.~(\ref{ham4}) and (\ref{H10m}). Beginning with the four-qubit system as shown in Fig.~\ref{fig:SingleQubit}, the logical states are
%\begin{align}
%\ket{0}_L &= \frac{-i}{\sqrt{2}}\ket{1000} + \frac{1}{\sqrt{2}}\ket{1011}\,,\nonumber\\
%\ket{1}_L &= \frac{i}{\sqrt{2}}\ket{1000} + \frac{1}{\sqrt{2}}\ket{1011}\,.
%\end{align}
\begin{equation}
    \begin{split}
        \ket{0}_L=&\frac{1}{2}\left(\ket{0101}+\ket{1010}\right)+ \frac{i}{2}\left(\ket{0110}+\ket{1001}\right)\\
        \ket{1}_L=&\frac{1}{2}\left(\ket{0100}+\ket{1011}\right)- \frac{i}{2}\left(\ket{0111}+\ket{1000}\right).
    \end{split}
\end{equation}

\begin{figure}
    [t]
    \[\Qcircuit @C=1em @R=1em @!R {
    \lstick{(a)\text{        }\ket{0}}& \gate{X}& \gate{Y}& \targ & \qw & \qw & \qw & \qw  \\
    \lstick{\ket{0}}& \gate{R_y\left(-\frac{\pi}{2}\right)} & \ctrl{-1} & \qw & \targ & \qw & \qw & \qw && \raisebox{-3em}{$\ket{0}_L$} \\
     \lstick{\ket{0}}& \gate{R_y\left(\frac{\pi}{2}\right)} & \qw & \ctrl{-2} & \ctrl{-1} & \ctrl{1}& \gate{X} & \qw   \\
     \lstick{\ket{0}}&  \qw & \qw & \qw & \qw & \targ & \qw & \qw 
    } \]

    \[\Qcircuit @C=1em @R=1em @!R {
    \lstick{(b)\text{        }\ket{0}}& \gate{R_x\left(\frac{\pi}{2}\right)}& \ctrl{3}& \targ & \ctrl{1} & \ctrl{3} & \qw & \qw  \\
    \lstick{\ket{0}}& \gate{X} & \qw & \qw & \targ & \qw & \qw & \qw && \raisebox{-3em}{$\ket{1}_L$} \\
     \lstick{\ket{0}}& \gate{H} & \qw & \ctrl{-2} & \qw & \qw& \qw & \qw   \\
     \lstick{\ket{0}}&  \qw & \targ & \qw & \qw & \targ & \qw & \qw 
    } \]

    \[\Qcircuit @C=1em @R=1em @!R {
    \lstick{(c)\text{        }\ket{0}}& \gate{R_x\left(\frac{\pi}{2}\right)}& \ctrl{3}& \targ & \ctrl{1} & \qw & \qw & \qw  \\
    \lstick{\ket{0}}& \gate{X} & \qw & \qw & \targ & \ctrl{1} & \qw & \qw && \raisebox{-3em}{$\ket{+}_L$} \\
     \lstick{\ket{0}}& \gate{H} & \qw & \ctrl{-2} & \qw & \control \qw & \qw & \qw   \\
     \lstick{\ket{0}}&  \qw & \targ & \qw & \qw & \qw & \qw & \qw 
    } \]

    \[\Qcircuit @C=1em @R=1em @!R {
    \lstick{(d)\text{        }\ket{0}}& \gate{R_x\left(\frac{\pi}{2}\right)}& \qw &  \ctrl{3}& \targ & \ctrl{1} & \qw & \qw & \qw  \\
    \lstick{\ket{0}}& \gate{X}& \gate{S}  & \qw & \qw & \targ & \ctrl{1} & \qw & \qw && \raisebox{-3em}{$\ket{i^+}_L$} \\
     \lstick{\ket{0}}& \gate{H}& \qw  & \qw & \ctrl{-2} & \gate{S^\dagger} & \control \qw & \ctrl{1}& \qw   \\
     \lstick{\ket{0}}&  \qw& \qw  & \targ & \gate{H} & \gate{S^\dagger} & \qw & \control \qw  & \qw 
    } \]
\caption{Circuits to initialize the required logical states (a) $\ket{0}_L$, (b) $\ket{1}_L$, (c) $\ket{+}_L$ and (d) $\ket{i^+}_L$, for process tomography for the 4-MZM, simulated topological qubit experiments.} 
\label{fig:Logical_Init_Y1}
\end{figure}
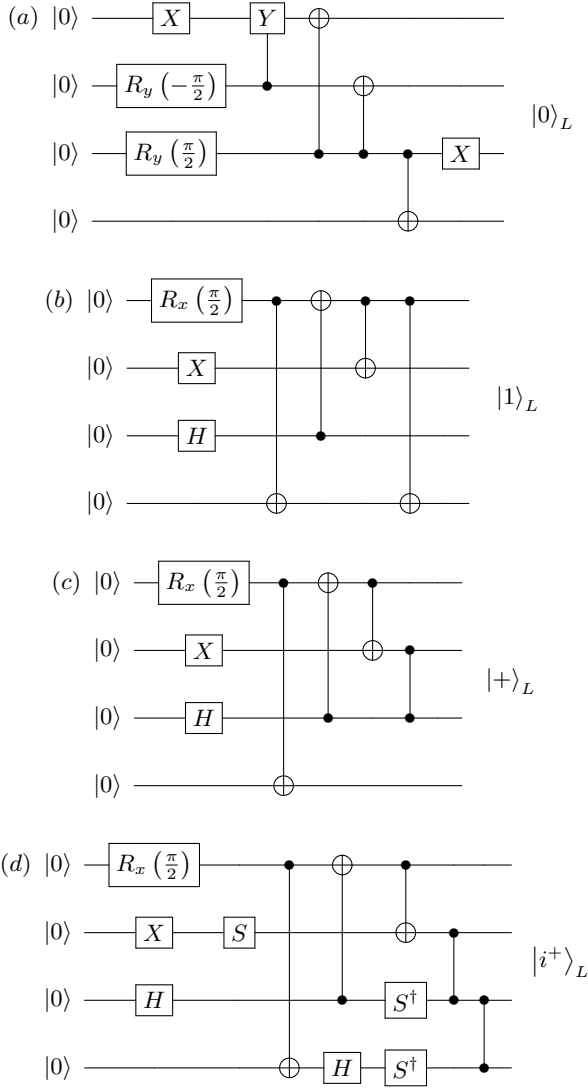

\noindent These states were selected as ground state eigenvalues of the integrals of motion operators given in Eq.~\ref{Wp}, and orthogonal eigenvalues of the logical basis operator $Z_L=-\sigma_2^z \sigma_3^z$. The circuits used to initialize these states are given in Fig.~\ref{fig:Logical_Init_Y1}, including circuits to prepare the $\ket{+}_L$ and $\ket{i^+}_L$ states, which are necessary for process tomography. To test the initialization and process fidelities of all MZM qubit simulations discussed, tomography is performed in the simulated qubit basis. To do so, along with the logical $Z_l$ axis of the simulated Bloch sphere, measurements on the system are done along the logical $X_l=\sigma_2^y\sigma_3^z$ and logical $Y_l=\sigma_2^x$. The derivation of these logical operators from the MZM Hamiltonian to the qubit simulation picture is given in App.~\ref{app:Kitaev_lattice}. Throughout this work the process fidelity is chosen as a metric of the quality for each simulate gate, as it accounts for the full simulated quantum channel. To derive the process fidelity, density matrices of the output state from the simulated gate are measured for the set of input states i.e. for a single qubit gate the set of $\{\ket{0}_L,\ket{1}_L,\ket{+}_L,\ket{i^+}_L\}$ is used. From these density matrices a Choi-matrix $\tilde\chi$ describing the quantum channel may be built and the fidelity of that channel relative to the lossless, noiseless case $\chi$ is given as 

\begin{equation}
    \mathcal{F}=\frac{\Tr[\chi^\dagger\tilde\chi]}{d^2}
\end{equation}

\noindent where $d$ is the size of the Hilbert space. The fidelities of initialization of the logical basis states on the 127 qubit \textit{ibm\_brisbane} Eagle r3 transmon device are given in Tab.~\ref{tab:TabFull_Brisbane}, as compared to classical simulations of the same device using the \textit{qiskit\_aer} package. This noisy intermediate scale quantum (NISQ) processor is used for all experimental simulations in this work.

A Trotter decomposition of the Hamiltonian in Eqs.~(\ref{ham4}) is employed to perform adiabatic clock face rotations on the simulated MZM qubits. The time evolution of Eqs.~(\ref{ham4}) is 

\begin{equation}
U_{CF}(t) = e^{-i t (\Delta_z(t)\sigma_0^z\sigma_1^z + \Delta_y(t)\sigma_0^y\sigma_2^y + \Delta_x(t)\sigma_0^x\sigma_3^x)}
\end{equation}

\noindent which for small $t\rightarrow \delta t$

\begin{equation}
\begin{split}
U_{CF}(\delta t) \approx e^{-i \delta t \Delta_x(\delta t)\sigma_0^x\sigma_3^x}
e^{-i \delta t \Delta_y(\delta t)\sigma_0^y\sigma_2^y }
e^{-i \delta t \Delta_z(\delta t)\sigma_0^z\sigma_1^z}&\\
 =R_{x_0x_3}(\delta t \Delta_x(\delta t))R_{y_0y_2}(\delta t \Delta_y(\delta t))R_{z_0z_1}(\delta t \Delta_z(\delta t))&
\end{split}
\end{equation}

\begin{figure}
    [t]
    \[\Qcircuit @C=1em @R=1em @!R {
    & \qw   & \multigate{1}{R_{ii}(|\tilde\Delta|\cos n \delta \theta)} & \qw & \qw & \qw &&& \raisebox{2.5em}{$\Pi_{n=0}^N$}\\
    & \qw &  \ghost{R_{ii}(|\tilde\Delta|\cos n \delta \theta)} & \qw & \multigate{1}{R_{jj}(|\tilde\Delta|\sin n \delta \theta)}& \qw &\\
    & \qw & \qw & \qw  & \ghost{R_{jj}(|\tilde\Delta|\sin n \delta \theta)} & \qw  \gategroup{1}{6}{3}{1}{1em}{(} \gategroup{1}{6}{3}{1}{1em}{)} 
    }\]
\caption{Circuit diagram of an example of the Trotter decompsition of the adiabatic tuning of the tunnel couplings of the 3 qubit the Hamiltonian $H_3 =|\tilde\Delta| \cos \theta \sigma_0^i\sigma_1^i+|\tilde\Delta| \sin \theta \sigma_1^i\sigma_2^i$ as $\theta$ is varied from $0$ to some angle $\phi$ in steps of $\delta\theta=\phi/N$.} 
\label{fig:Trotter_circ_2}
\end{figure}
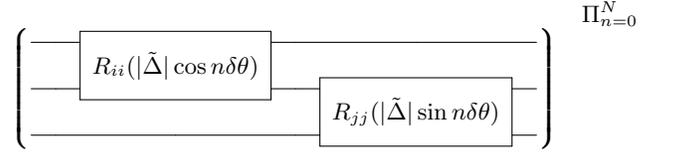

\begin{figure}
	[b] 
	\raggedright
	\includegraphics[width=0.48\linewidth]{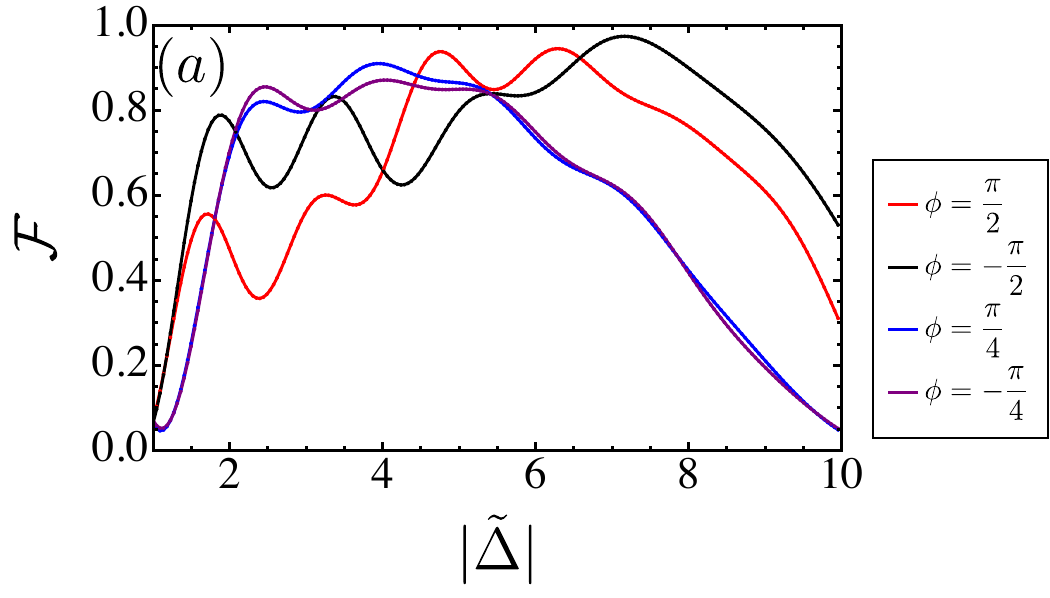} 
	\includegraphics[width=0.48\linewidth]{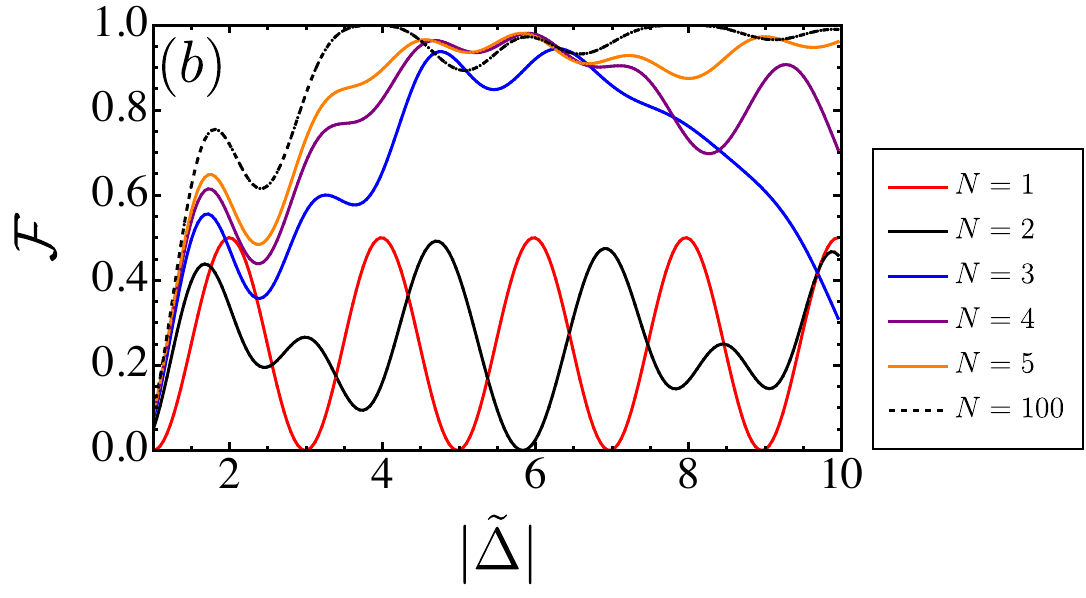} 
	\caption{Optimizing the Trotterized form the the adiabatic evolution of exchange couplings about a MZM Y-junction. (a) Process fidelity from noiseless state vector simulation of a constant $\delta \theta=\phi/3$ step Trotter decomposition for the relevant $\phi$ as a function of $|\tilde\Delta|$. (b) Process fidelity from noiseless state vector simulation of $\delta \theta=\phi/N$ step Trotter decomposition at different $N$ as a function of $|\tilde\Delta|$.}
\label{fig:N_vs_F_1Y} \end{figure}

\noindent where $R_{ij}(\phi)=\exp\left(-i\phi\sigma^i\sigma^j/2\right)$. Taking the coupling parameters as $\vec\Delta=|\Delta|(\sin\theta\cos\phi,\sin\theta\sin\phi,\cos\theta)$, the total evolution around the clock face made up of three segments or paths $\{\theta,\phi\}: \{0,0\}\rightarrow\{\pi/2,0\}$, $\{\pi/2,0\}\rightarrow\{\pi/2,\phi\}$ and $\{\pi/2,\phi\}\rightarrow\{0,0\}$. For each path, a Trotter decomposition of $N$ steps is applied to the simulated qubit state

\begin{equation}
    U_{T} = \Pi_{n=0}^N R_{i_0i_i}(|\tilde\Delta| \cos n \delta\theta)R_{j_0j_j}(|\tilde\Delta| \sin n \delta\theta)
\end{equation}

\noindent where $\delta\theta = \tilde\theta/N$, $\tilde\theta =\pi/2$ or $\tilde\theta =\phi$ depending on the angle varied along the evolved path and $|\tilde\Delta|=\delta t |\Delta|$ is the global exchange constant of the simulation. This constant can be varied to optimize the process fidelities of the simulated evolutions. The effect of $|\tilde\Delta|$ is shown in Fig.~\ref{fig:N_vs_F_1Y}, along side the effect of varying $N$. For consistency, for each path of the total evolution, $N$ is chosen separately to ensure that $\delta\theta$ is constant throughout. For the case where the magnitude of maximum $|\phi|=\pi/2$ of the chosen evolution is equal to the maximum $\theta=\pi/2$, $N$ is constant for each path of the total evolution. However, for the cases when the magnitude of maximum $|\phi|\neq \pi/2$, the condition $N_{\{0,0\}\rightarrow\{\pi/2,0\}}=N_{\{\pi/2,\phi\}\rightarrow\{0,0\}}=\pi N_{\{\pi/2,0\}\rightarrow\{\pi/2,\phi\}}/2|\phi|$ is obeyed.

\begin{table}
    [t]
    \centering
    \begin{tabular}{|l||c|c|c|c|}
        \hline
         Operation & Simulation  &  Experiment &  Av. Depth &  $|\tilde\Delta|$ \\
        \hline
        \hline
        $I$ & $91.52\pm1.31\%$ & $85.12\pm1.42\%$ & 21 & -\\
        $S$ & $83.78\pm1.50 \%$ & $84.50\pm9.44 \%$ & 137 & 6.3 \\
        $T$ & $70.29\pm1.30\%$ & $41.90\pm1.60\%$ & 302 & 4.2\\
        $S^{\dagger}$ & $84.71\pm1.49\%$ & $74.48\pm1.47\%$ & 182 & 7.0\\
        $T^{\dagger}$ & $73.35\pm1.27 \%$ & $53.44\pm5.36 \%$ & 323 & 4.0 \\
        \hline
    \end{tabular}
    \caption{Initialization and 4-MZM topological qubit simulated process fidelity simulation results, experimental results, average circuit depth and associated $|\tilde\Delta|$ as performed on the \textit{ibm\_brisbane} quantum processor. Here, each experiment consists of $2^{13}$ shots.}
    \label{tab:TabFull_Brisbane}
\end{table}

\begin{figure}
    [b]
    \includegraphics[width=0.98\linewidth]{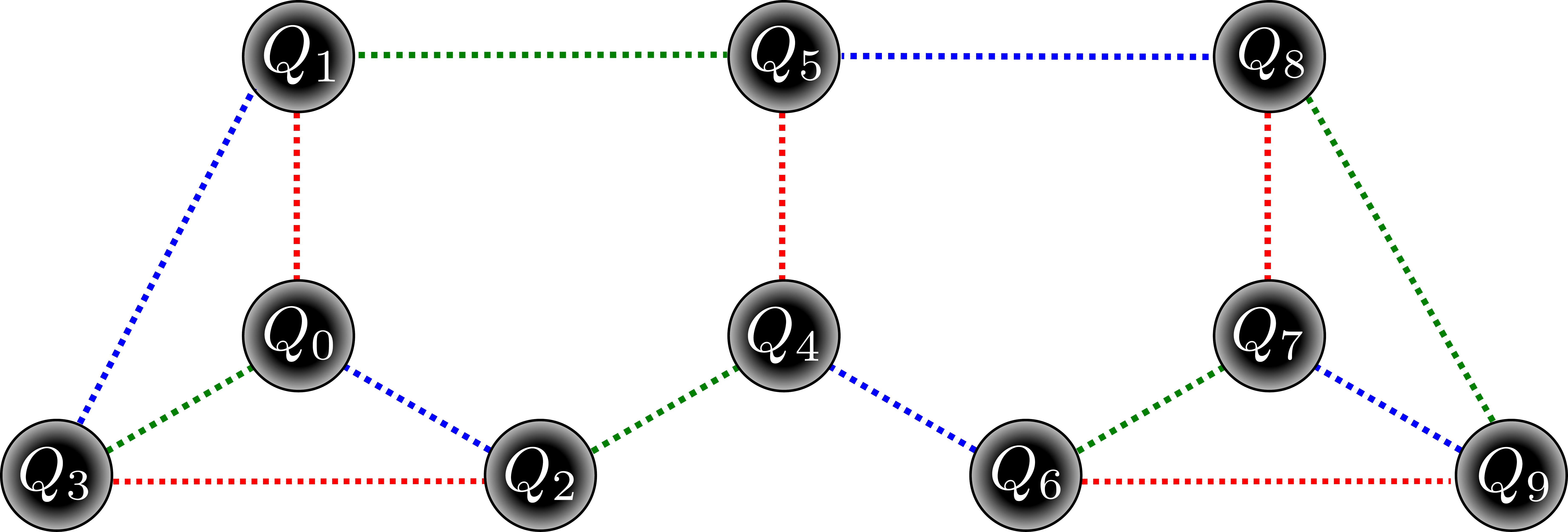}
    \caption{Qubit analogue of the 10-MZM, 2 simulated topological qubits with closed Kiteav lattice like connectivity. The colored dashed lines indicate the nature of the coupling between neighboring qubits in the simulation Hamiltonian, and thus the measurements used to perform operations. Red indicates $\sigma_i^z \sigma_j^z=(i \gamma_i^z\gamma_i)(i \gamma_j^z\gamma_j)$, blue indicates $\sigma_i^y \sigma_j^y=(i \gamma_i^y\gamma_i)(i \gamma_j^y\gamma_j)$ and green indicates $\sigma_i^x \sigma_j^x=(i \gamma_i^x\gamma_i)(i \gamma_j^x\gamma_j)$.}
    \label{fig:DoubleQubit}
\end{figure}

The process tomography results for the simulation of the adiabatic MZM braiding is given in Tab.~\ref{tab:TabFull_Brisbane}. For all processes $S^{(\dagger)}=R_z\left(\phi=\pm\pi/2\right)$ and $T^{(\dagger)}=R_z\left(\phi=\pm\pi/4\right)$ a constant time step of $\delta\theta=\phi/3$ was chosen. Therefore, the total number of Trotter steps for the $S^{(\dagger)}$ gates is $9$, $3$ per path of the total evolution, whilst for the $T^{(\dagger)}$ gates the total number of Trotter steps is $15$ since $N_{\{0,0\}\rightarrow\{\pi/2,0\}}=N_{\{\pi/2,\phi\}\rightarrow\{0,0\}}=6$. This is reflected in the depth of the corresponding circuits. Accordingly, the observed process fidelities of the $S^{(\dagger)}$ gates are significantly higher than those of the $T^{(\dagger)}$ gates, with up to $84.50\pm9.44\%$ fidelity. Therefore, this initial set of proof-of-concept simulations of adiabatic braiding operations on MZM qubits is successful, despite scope for improvements by way of Trotter step and $\vec\Delta$ optimization.

% \begin{figure}
%     [t]
%     \[\Qcircuit @C=0.41em @R=0.6em @!R {
%     & \qw &  \multigate{1}{R_{zz}(|\Delta|)}& \qw  & \multigate{1}{R_{zz}(|\Delta|\cos \delta \theta_n)} & \qw & \qw &&&&& \raisebox{3em}{$N-2$}\\
%     & \qw &  \ghost{R_{zz}(|\Delta|)} & \qw & \ghost{R_{zz}(|\Delta|\cos \delta \theta_n)} & \multigate{1}{R_{yy}(|\Delta|\sin \delta \theta_n)}& \qw &\rstick{\dots}\\
%     & \qw & \qw & \qw  & \qw & \ghost{R_{yy}(|\Delta|\sin \delta \theta_n)} & \qw &
%     }\]

%     \[\Qcircuit @C=0.41em @R=0.6em @!R {
%     & \qw & \qw & \qw & \qw\\
%    \lstick{\dots} & \qw & \multigate{1}{R_{yy}(|\Delta|)} & \qw & \qw \\
%     & \qw & \ghost{R_{yy}(|\Delta|)} & \qw & \qw \\
%     }\]
% \caption{.} 
% \label{fig:Trotter_circ_1}
% \end{figure}

% \sh{How are these generated? Could we not just use something very simple like $\ket{1000}$ and $\ket{0000}$. Also, one needs ot make connection with the previous section? Which qubits are which qubits? Is it really like the first number corresponds qubit zero?}
With the same tools, the aforementioned entangling braiding operations can be simulated. For the 2-topological qubit case, a 10-MZM system mapped to 10 physical qubits with Kitaev lattice like connectivity as shown in Fig.~\ref{fig:DoubleQubit}, is simulated. By deriving a set of integrals of motion operators, simulated logical qubit operators and energy state operators, similarly to the single simulated MZM qubit case and discussed completely in App.~\ref{app:Kitaev_lattice}, the following simulated logical states were chosen

\begin{widetext}
\begin{equation}
    \begin{split}
        \ket{0}^{q_0}_L=&\frac{\sqrt{2}}{4}\left(\ket{010101}+\ket{010110}+\ket{101001}+\ket{101010}\right)+\frac{i\sqrt{2}}{4}\left(\ket{011001}+\ket{011010}+\ket{100101}+\ket{100110}\right)\\
        \ket{1}^{q_0}_L=&-\frac{\sqrt{2}}{4}\left(\ket{011101}-\ket{011110}+\ket{100001}-\ket{100010}\right)-\frac{i\sqrt{2}}{4}\left(\ket{010001}-\ket{010010}+\ket{101101}-\ket{101110}\right)\\
        \ket{0}^{q_1}_L=&\frac{\sqrt{2}(i+1)}{4}\left(\ket{0110}+\ket{1001}\right)- \frac{\sqrt{2}(i-1)}{4}\left(\ket{0101}+\ket{1010}\right)\\
        \ket{1}^{q_1}_L=&\frac{\sqrt{2}(i-1)}{4}\left(\ket{0111}+\ket{1000}\right)- \frac{\sqrt{2}(i+1)}{4}\left(\ket{0100}+\ket{1011}\right).\\
    \end{split}
\end{equation}
\end{widetext}

\noindent  Circuits were derived to initialise all logical states needed for tomography in this basis, which are given in Figs.~\ref{fig:Logical_Init_Y2_Q0} and~\ref{fig:Logical_Init_Y2_Q1} in App.~\ref{app:10_init_Circuits}. These chosen basis states have the following logical operators, derived from the same set of operators, which gives the measurement axes for tomography

\begin{equation}
    \begin{split}
        X_L^{q_0} &= i\sigma_2^x \sigma_5^z\\
        Y_L^{q_0} &= -i\sigma_2^y \sigma_3^z \sigma_5^z\\
        Z_L^{q_0}&= -i\sigma_2^z \sigma_3^z
    \end{split}
    \quad\quad
    \begin{split}
        X_L^{q_1} &= -i\sigma_6^y\\
        Y_L^{q_1} &= -i\sigma_6^x \sigma_9^z\\
        Z_L^{q_1} &= -i\sigma_6^z \sigma_9^z.
    \end{split}
\end{equation}

With the logical states initialized and measurement axes derived, the same Trotterized adiabatic evolution of the exchange couplings is employed to perform operations. As discussed in Sec.~\ref{yjunc}, different braiding operations can be performed by adiabatic evolution of the exchange couplings around each Y-junction or clock face. Here, we focus our simulations of evolutions around the central clock face, as to simulate the entangling $R_{xx}\left(\phi\right)$ operation, as the braiding operations performed by evolutions around the left- and right-most clock faces are equivalent to the operations performed on the single topological qubit already discussed. Despite the different operation being simulated in the qubit space, the Hamiltonian being Trotterized to achieve entanglement is the same as in the single topological qubit simulations. As such, the optimization conditions of the parameters $|\tilde\Delta|$ and $N$ depicted in Fig.~\ref{fig:N_vs_F_1Y} apply here. 

\begin{table}
    [b]
    \centering
    \begin{tabular}{|l||c|c|c|c|}
        \hline
         Operation & Simulation  &  Experiment &  Av. Depth &  $|\tilde\Delta|$ \\
        \hline
        \hline
        $I$ & $81.26\pm1.01\%$ & $85.97\pm0.95\%$ & 34 & -\\
        $R_{xx}\left(\frac{\pi}{2}\right)$ & $63.77\pm1.11\%$ & $47.11\pm1.42\%$ & 149 & 6.3\\
        $R_{xx}\left(-\frac{\pi}{2}\right)$ & $66.72\pm0.87\%$ & $47.73\pm1.15\%$ & 167 & 7.2\\
        \hline
    \end{tabular}
    \caption{Initialization and $R_{xx}\left(\pm\pi/2\right)$ simulated process fidelity simulation results, experimental results, average circuit depth and associated $|\tilde\Delta|$ as performed on the \textit{ibm\_brisbane} quantum processor. Here, each experiment consists of $2^{13}$ shots.}
    \label{tab:Rxx_Tab}
\end{table}

The results of the initialization fidelity, and the $R_{xx}\left(\pm\pi/2\right)$ process fidelities are given in Tab.~\ref{tab:Rxx_Tab}. As expected, the circuit depths of the $R_{xx}\left(\pm\pi/2\right)$ gates are similar to that of the $S^{(\dagger)}$ gates in Tab.~\ref{tab:TabFull_Brisbane}, however the observed process fidelities, both experimental and in simulation, are significantly reduced. This is due to the increased susceptibility larger physical and simulated logical qubit spaces to decoherence during the applied circuits, despite the application of an $XY-4$ dynamical decoupling sequence to all reported experimental process fidelities. This is evident in the bias in the noise of the observed density matrices on the real diagonal elements, as is shown in App.~\ref{app:Sim_Data} (Fig.~\ref{fig:Hinton_RX}). Although the process fidelities of the entangling braiding operations top out at $47.73\pm1.15\%$, the results of each of the state fidelity experiments, from which the process fidelity is calculated, paints a better proof-of-concept picture with fidelities up to $72.77\pm0.77\%$. These results are shown in Tab.~\ref{tab:TabFull_Brisbane_10qubit_PlusPi2} and Tab.~\ref{tab:TabFull_Brisbane_10qubit_MinusPi2} in App.~\ref{app:Sim_Data}.

\begin{comment}
$\gamma_i$ and $\gamma_i^\alpha$ are Majoranas satisfying $(\gamma^\alpha_i)^2=(\gamma_i)^2$ and $\{\gamma^\alpha_i,\gamma^\beta_j\}=\{\gamma^\alpha_i,\gamma_j\}=\{\gamma_i,\gamma_j\}=0$ for $i\neq j$ or $\alpha\neq\beta$. Inserting this into Eq.~(\ref{ham4}), we obtain
\begin{align}
    H &= i[\Delta_z(\gamma_0 \gamma_1) + \Delta_y(\gamma_0\gamma_2)  + \Delta_x(\gamma_0\gamma_3) ] \nonumber \\
     &= i\gamma_0(\vec{\Delta} \cdot \vec{\gamma})\,, \nonumber
\end{align}
where we have defined $\vec{\gamma} = (\gamma_1, \gamma_2, \gamma_3)$ and $\vec{\Delta} = (\Delta_x, \Delta_y, \Delta_z)$. Because $[\gamma^z_1\gamma_2^z,H]=[\gamma^y_1\gamma_3^y,H]=[\gamma^x_1\gamma_4^x,H]=0$, i.e. these products of Majoranas are conserved quantities, we are free to fix $i\gamma^z_1\gamma_2^z=i\gamma^y_1\gamma_3^y=i\gamma^x_1\gamma_4^x=1$.

Within this section: 10 qubit case

In the next section: simulating in mathematica and the projection

In the following section: simulating on a QC

In the discussion: comments on error correction perhaps why it works and relation to majorana latices, measurement based quantum computing in these systems, 
 
This encoding and operation can be likewise described as qubit operations. 

First, note that the Hilbert space encoded by the qubits is four times larger than 
 
 and we choose to start in the low energy subspace without loss of generality; 

These four physical qubits encode one logical qubit whose states are defined by the eigenvalues of $i\gamma_3\gamma_4$. 

\end{comment}

\section{Discussion and Conclusion}
\label{disc}

In this work, the simulation of adiabatic braiding operations of MZM qubits in a Kitaev-lattice like structure was demonstrated on a NISQ processor. A rigorous formalism for the qubitization of the MZM Y-junctions allowed for the derivation of logical basis used to simulate the topological qubit, as well as a set of measurement axes with which tomography in the logical basis was performed. A Trotterized form of the MZM Y-junction Hamiltonian allowed for the simulation of the adiabatic braiding of MZMs on the Y-junction, resulting in observable unitary operations in the simulated qubit basis. This same Trotterized evolution was used to demonstrate $S^{(\dagger)}$ braiding operations with up to $84.50\pm9.44\%$ process fidelity and $T^{(\dagger)}$ partial braiding operations with up to $53.44\pm5.36\%$ process fidelity on a single simulated MZM Y-junction. The drop in process fidelity of the $T^{(\dagger)}$ can be attributed to the approximately doubling of the depth of the circuits investigated. The same Trotterized form of the simulated adiabatic braiding of MZMs was also shown to perform an entangling $R_{xx}(\pm \pi/2)$ operation between two simulated topological qubits on a 10-MZM lattice with process fidelities up to $47.73\pm1.15\%$. Here, operational fidelity was primarily hampered by dephasing errors on the extended physical and simulated qubit spaces compared to the single simulated qubit operations. Overall, the simulations demonstrate the viability of simulating the braiding of MZMs with NISQ hardware. Throughout the simulations performed in this work, a constant maximum exchange coupling parameter is assumed for each arm of the Y-junctions tested. This assumption, although consistent to the original proposal\cite{karzig2016universal}, is ultimately what leads to the large circuit depths limiting the process fidelities of the $T^{(\dagger)}$ partial braiding operations. By relaxing this assumption, and allowing for an asymmetric Y-junction, the Trotterized Hamiltonian evolution may be further optimized.

Beyond the simulation of the adiabatic exchange of MZMs to demonstrate an initial, restrictive gate set, the simulation tools demonstrated here could serve as platform to demonstrate gate in alternative MZM qubit encoding. For example, in the 10-MZM picture discussed in this work, a universal gate set may be implemented on an encoded a single qubit within two-qubit system (i.e. $\ket{0}=\ket{0}_L^{q_0}\ket{1}_L^{q_1}$ and $\ket{1}=\ket{1}_L^{q_0}\ket{0}_L^{q_1}$) such that the operations demonstrated here could be combined to offer full single qubit control. It is important to note however, that although the adiabatic exchange gates discussed in this work could be extended to offer universal control of a MZM qubit, the gates themselves are geometric and therefore are not topologically protected as their are susceptible to errors in dynamical phases gather from imperfect exchange. To minimize such errors, similar operations have been proposed by sequences of charge-parity measurements\cite{karzig2017scalable} instead of adiabatic tuning of exchange. Such operations can equally be simulated by sequences of the appropriate parity checks\cite{brooks2025measurement}, without the need for the approximate Trotterization of the simulated Hamiltonian. Such parity check sequences and logical state encoding is analogous to a form of quantum error correcting (QEC) codes known as Floquet codes\cite{hastings2021dynamically, dua2024engineering, zhang2023x, davydova2023floquet}, and so simulation of such operations and codes could demonstrate MZM qubits as a natural companion to such QEC codes. Finally, the qubitization of Kitaev-like connected structures discussed could open the door to the simulation of other topologically non-trivial condensed matter phenomena on a Kitaev lattice such as topological edge currents\cite{kitaev2006anyons,thakurathi2014majorana,mcginley2017robustness, self2017conformal,busnaina2024quantum,bespalova2021quantum}.

%Furthermore, similar braiding operations have been proposed by sequences of charge parity measurements\cite{karzig2017scalable} instead of adiabatic tuning of exchange. Such operations could equally be simulated by sequences of the appropriate parity checks, without the need for the approximate Trotterization of the simulated Hamiltonian

\section{Acknowledgments}

We acknowledge helpful discussions with L. Giovia, U. G\"ung\"ord\"u, and Y. Yanay. This work is supported by the U.S. Department of Energy, Office of Science, Basic Energy Sciences under Award No. DE-SC0022089. 

\bibliography{Bibliography}

\begin{widetext}
\pagebreak
\appendix

\section{Qubits to Majoranas}
\label{app:Kitaev_lattice}

While Eq.~(\ref{Hmaj}) is a portion of the Kitaev lattice on a plane, it is instructive to consider the Kitaev tetrad on a torus wherein the Hamiltonian is
\begin{equation}
    H_{T^2}= \Delta_z\sigma_0^z\sigma_1^z + \Delta_y\sigma_0^y\sigma_2^y + \Delta_x\sigma_0^x\sigma_3^x+\bar\Delta_z\sigma_2^z\sigma_3^z + \bar\Delta_y\sigma_1^y\sigma_3^y + \bar\Delta_x\sigma_1^x\sigma_2^x\,.
    \label{HT2}
\end{equation}
One can show that there exists integrals of motion, 
\begin{align}
    W_1&=\sigma^z_0\sigma^x_2\sigma^y_3\nonumber\,,\\
    W_2&=\sigma^y_0\sigma^x_1\sigma^z_3\nonumber\,,\\
    W_3&=\sigma^x_0\sigma^y_1\sigma^z_2\,,
    \label{Wp}
\end{align}
which commute with $H_{T^2}$ and have eignevalues $\pm1$. Our logical qubit will be idle when $\Delta_x=\Delta_y=\bar\Delta_x=\bar\Delta_y=\bar\Delta_z=0$. The low and high energy states will be characterized by eigenvalues of the Hamiltonian at that point in parameter space, i.e. the operator $h=\sigma_0^z\sigma_1^z$, and the degenerate states defining the logical basis of the qubit correspond to eigenvalues of the operator $n=\sigma_2^z\sigma^z_3$. Although the integrals of motion commute with $h$ and $n$, their eigenvalues are not independent, $W_2W_3= h n$. %We will henceforth work in the low-energy space which, because we our parameters our changed adiabatically, can be defined by the eigenvalue of the $h$ to be $-1$ when idle. Moreover, 
We choose to work in the subspace in which $W_1$ and $W_2$ have eigenvalues of $-1$ without loss of generality; the eigenvalue of $W_3$ is then determined by the qubit state and energy subspace.

The Pauli matrices $\sigma^\alpha_j$ can be represented by the operators $\tilde\sigma^\alpha_j=i\gamma_j^\alpha \gamma_j$ in an extended space. The physical subspace is recovered by restricting the extended space to the set of states on which $D_j=-i\tilde\sigma^x_j\tilde\sigma^y_j\tilde\sigma^z_j=\gamma_j^x\gamma_j^y\gamma_j^z\gamma_j$ acts as the identity. In the extended space, Eq.~(\ref{HT2}), Eq.~(\ref{Wp}), $h$ and $n$ become
\begin{align}
     \tilde H_{T^2}&= -i(\Delta_z \hat u^z_{01}\gamma_0\gamma_1 + \Delta_y\hat u^y_{02}\gamma_0\gamma_2 + \Delta_x\hat u^x_{03}\gamma_0\gamma_3+\bar\Delta_z\hat u^z_{23}\gamma_2\gamma_3 + \bar\Delta_y\hat u^y_{13}\gamma_1\gamma_3 + \bar\Delta_x\hat u^x_{12}\gamma_1\gamma_2)\,,\nonumber\\
   \tilde W_1&=-\hat u^x_{03}\hat u^y_{02}\hat u^z_{23}\,,\nonumber\\
    \tilde W_2&=\hat u^x_{03}\hat u^y_{13}\hat u^z_{01}\,,\nonumber\\
    \tilde W_3&=-\hat u^z_{01}\hat u^x_{12}\hat u^y_{02}\,,\nonumber\\
    \tilde h&=-i\hat u^z_{01}\gamma_0\gamma_1\,,\nonumber\\
    \tilde n&=-i\hat u^z_{23}\gamma_2\gamma_3\,,
    \label{HT2_ex}
\end{align}
where $\hat u^\alpha_{ij}=i\gamma^\alpha_i\gamma^\alpha_j$ are integrals of motion, i.e. commute amongst eachother, with $\tilde H_{T^2}$, and with $\tilde h$ and $\tilde n$\cite{kitaev2006anyons}. It is consistent to take $-\hat u^z_{01}=-\hat u^y_{02}=-\hat u^x_{03}=\hat u^z_{23}=\hat u^y_{13}=1$ wherein $\tilde W_1=\tilde W_2=-1$ and, upon setting $\bar\Delta_z=\bar\Delta_y=\bar\Delta_x=0$, Eq.~(\ref{HT2_ex}) reduces to
\begin{align}
     \tilde H_{T^2}\rightarrow H&= i(\Delta_z\gamma_0\gamma_1 + \Delta_y\gamma_0\gamma_2 + \Delta_x\gamma_0\gamma_3)\,,\nonumber\\
    \tilde h\rightarrow h&=i\gamma_0\gamma_1\,,\nonumber\\
    \tilde n\rightarrow n&=-i\gamma_2\gamma_3\,.
\end{align}
As it will be convenient for tomography, we note that the logical Pauli operators are $Y_L=\sigma^y_2\sigma^z_3\rightarrow\gamma_3\gamma_2^x \hat u_{23}^z\rightarrow\gamma_3$, $X_L=\sigma^x_2\rightarrow \gamma_2\gamma_2^x\rightarrow\gamma_2$ and $Z_L=\sigma_2\sigma_3\rightarrow\gamma_2\gamma_3$ where the simplification of $Y_L$ and $X_L$ are the result of $\gamma^x_2$ commuting with the operators in Eq.~(\ref{HT2_ex}) and we have ignored factors of $-1$ and $i$. 

An analogous lattice of ten spins can be written as
\begin{align}
    H_{10,T^2}&= \Delta_z\sigma_0^z\sigma_1^z + \Delta_y\sigma_0^y\sigma_2^y + \Delta_x\sigma_0^x\sigma_3^x+\bar\Delta_z\sigma_2^z\sigma_3^z + \bar\Delta_y\sigma_1^y\sigma_3^y + \bar\Delta_x\sigma_1^x\sigma_5^x\nonumber\\
    &+\Delta_z'\sigma_{0'}^z\sigma_{1'}^z + \Delta_y'\sigma_{0'}^y\sigma_{2'}^y + \Delta_x'\sigma_{0'}^x\sigma_{3'}^x+\bar\Delta_z'\sigma_{2'}^z\sigma_{3'}^z + \bar\Delta_y'\sigma_{1'}^y\sigma_{5}^y + \bar\Delta_x'\sigma_{1'}^x\sigma_{2'}^x\nonumber\\
    &+\Lambda_z\sigma_4^z\sigma_5^z +\Lambda_y\sigma_4^y\sigma_{3'}^y+\Lambda_x\sigma_2^x\sigma_4^x\,.
    \label{H10}
\end{align}
In addition to $W_1$ and $W_2$, there are four more independent integrals of motion,
\begin{align}
    W_{4}&=\sigma^z_{0'}\sigma^x_{2'}\sigma^y_{3'}\\\nonumber
    W_{5}&=\sigma^x_{0'}\sigma^y_{1'}\sigma^z_{2'}\\\nonumber
    W_{6}&=\sigma^x_0\sigma^y_1\sigma^z_2\sigma_4^y\sigma_5^y\\\nonumber
    W_{7}&=\sigma_4^x\sigma_5^x\sigma^y_{0'}\sigma^x_{1'}\sigma^z_{3'}\,.
\end{align}
$W_3$ does not commute with $H_{10,T^2}$. We treat these ten physical qubits as two logical qubits, encoded in qubits $\{0,1,2,3\}$ and $\{0',1',2',3'\}$, interfaced by qubits 4 and 5. The system is idle when only $\Delta_z,\Delta_z'$ and $\Lambda_z$ are nonzero wherein the spectrum is defined by the eigenstates of h, $h'=\sigma^z_{0'}\sigma^z_{1'}$, and $h_a=\sigma^z_4\sigma^z_5$. The logical qubit states are defined by the eigenvalues of $n$ and $n'=\sigma_{2'}^z\sigma_{3'}^z$. While $n,n',h$ and $h'$ commute with all the integrals of motion of the ten qubit system, $W_7\sim nn'hh'h_aW_1W_6'$ and their eigenvalues are not independent. Without loss of generality, we choose $W_j=-1$ for $j=1,2,4,5,6$. In the extended space,
\begin{align}
    \tilde H_{10,T^2}&=-i(\Delta_z \hat u^z_{01}\gamma_0\gamma_1 + \Delta_y\hat u^y_{02}\gamma_0\gamma_2 + \Delta_x\hat u^x_{03}\gamma_0\gamma_3+\bar\Delta_z\hat u^z_{23}\gamma_2\gamma_3 + \bar\Delta_y\hat u^y_{13}\gamma_1\gamma_3 + \bar\Delta_x\hat u^x_{15}\gamma_1\gamma_5\nonumber\\
    &+\Delta_z' \hat u^z_{0'1'}\gamma_0'\gamma_1' + \Delta_y'\hat u^y_{0'2'}\gamma_0'\gamma_2' + \Delta_x'\hat u^x_{0'3'}\gamma_0'\gamma_3'+\bar\Delta_z'\hat u^z_{2'3'}\gamma_2'\gamma_3' + \bar\Delta_y'\hat u^y_{1'5}\gamma_1'\gamma_5 + \bar\Delta_x'\hat u^x_{1'2'}\gamma_1'\gamma_2'\nonumber\\
    &+\Lambda_z \hat u^z_{45}\gamma_4\gamma_5 +\Lambda_y \hat u^y_{43'}\gamma_4\gamma_{3'}+\Lambda_x \hat u^x_{42}\gamma_4\gamma_2\nonumber\,,\\
    \tilde W_4&=-\hat u^x_{0'3'}\hat u^y_{0'2'}\hat u^z_{2'3'}\,,\nonumber\\
    \tilde W_5&=-\hat u^z_{0'1'}\hat u^x_{1'2'}\hat u^y_{0'2'}\,,\nonumber\\
    \tilde W_6&=-\hat u^z_{01}\hat u^x_{15}\hat u^y_{02}\hat u_{42}^x \hat u_{45}^z\,,\nonumber\\
    \tilde W_7&=\hat u_{43'}^y u_{45}^z\hat u^x_{0'3'}\hat u^y_{1'5}\hat u^z_{0'1'}\,,\nonumber\\
    \tilde h'&=-i\hat u^z_{0'1'}\gamma_0'\gamma_1'\,,\nonumber\\
    \tilde n'&=-i\hat u^z_{2'3'}\gamma_2'\gamma_3'\,,\nonumber\\
    \tilde h_a&=-i\hat u^z_{45}\gamma_4\gamma_5\,,
    \label{H10_ex}
\end{align}
where we have used the redundant notation $\gamma_{j'}=\gamma_j'$. Again, because $u^\alpha_{ij}$ are integrals of motion, it is consistent with the choice of $W_j$ to take $\hat u^z_{01}=\hat u^y_{02}=\hat u^x_{03}=-\hat u^z_{23}=\hat u^y_{13}=\hat u^z_{0'1'}=\hat u^y_{0'2'}=\hat u^x_{0'3'}=-\hat u^z_{2'3'}=-\hat u^x_{1'2'}=-\hat u^y_{1'3'}=\hat u^z_{45}=\hat u^y_{43'}=\hat u^x_{42}=-\hat u^x_{15}=-1$. Upon setting $\bar\Delta_\alpha=\bar\Delta_\alpha'=0$,
\begin{align}
    \tilde H_{10,T^2}\rightarrow H_{10}&=i(\Delta_z \gamma_0\gamma_1 + \Delta_y\gamma_0\gamma_2 + \Delta_x\gamma_0\gamma_3
    +\Delta_z' \gamma_0'\gamma_1' + \Delta_y'\gamma_0'\gamma_2' + \Delta_x'\gamma_0'\gamma_3'\nonumber\\
&+\Lambda_z \gamma_4\gamma_5 +\Lambda_y \gamma_4\gamma_{3'}+\Lambda_x \gamma_4\gamma_2)\nonumber\,,\\
    \tilde h'\rightarrow h'&=i\gamma_0'\gamma_1'\,,\nonumber\\
    \tilde n'\rightarrow n'&=-i\gamma_2'\gamma_3'\,,\nonumber\\
    \tilde h_a\rightarrow h_a&=i\gamma_4\gamma_5\,.
    \label{H10_ex}
\end{align}
Thus, replacing $\gamma_4$ and $\gamma_5$ with $\zeta_0$ and $\zeta_1$, respectively, we recover Eq.~(\ref{Hmaj10}) in the main text. Notably, $n$ and $n'$ differ in sign from the usual definition in the usual Majorana qubit definition. Lastly, we note the logical Pauli operators are $Y_L=\sigma_2^y\sigma_3^z\sigma^z_5\rightarrow \gamma_3\gamma_5^y\hat u_{23}^z\hat u_{25}^x\rightarrow\gamma_3$, $X_L=\sigma_0^z\sigma_1^z\sigma_2^x\sigma^z_5\rightarrow\gamma_2$, $Z_L=\sigma_0^z\sigma_1^z\rightarrow\gamma_2\gamma_3$, $Y_L'=\sigma^y_{2'}\sigma^z_{3'}\rightarrow\gamma_{3'}$, $X_L'=\sigma^x_{2'}\rightarrow\gamma_{2'}$ and $Z_L'=\sigma^z_{0'}\sigma^z_{1'}\rightarrow\gamma_{2'}\gamma_{3'}
$.

\section{Effective Hamiltonian and Gauge potentials}
\label{Heffandgauge}
Because $n$, $h$, $W_1$, and $W_2$ commute with each other, we can define a complete orthonormal set of states, $|\psi_{m_1m_2m_3m_4}\rangle$, which are eigenstates of $n$, $h$, $W_1$, and $W_2$ with eigenvalues, $m_1$, $m_2$, $m_3$, and $m_4$, respectively. Because $W_1$ and $W_2$ also commute with the $H$, the $16\times16$ can be block diagonalized into four $4\times4$ identical blocks with different values $W_1$ and $W_2$. Without loss of generality we focus on the block with $W_1=W_2=-1$ but our analysis proceeds identically for any combination of values.

We use the states $|\psi_{11-1-1}\rangle$,$|\psi_{1-1-1-1}\rangle$,$|\psi_{-11-1-1}\rangle$, and $|\psi_{-1-1-1-1}\rangle$ as a basis. Upon defining the effective Pauli matrices,
\begin{align}
\tau^x=&|\psi_{-1-1-1-1}\rangle\langle\psi_{-11-1-1}|+|\psi_{1-1-1-1}\rangle\langle\psi_{11-1-1}|+\textrm{H.c.}\,,\nonumber\\
\tau^z=&|\psi_{-1-1-1-1}\rangle\langle\psi_{-1-1-1-1}|+|\psi_{-11-1-1}\rangle\langle\psi_{-11-1-1}|+|\psi_{1-1-1-1}\rangle\langle\psi_{1-1-1-1}|+|\psi_{11-1-1}\rangle\langle\psi_{11-1-1}|\,,\nonumber\\
\eta^x=&|\psi_{-1-1-1-1}\rangle\langle\psi_{1-1-1-1}|+|\psi_{-11-1-1}\rangle\langle\psi_{11-1-1}|+\textrm{H.c.}\,,\nonumber\\
\eta^y=&-i(|\psi_{-1-1-1-1}\rangle\langle\psi_{1-1-1-1}|+|\psi_{-11-1-1}\rangle\langle\psi_{11-1-1}|)+\textrm{H.c.}\,,\nonumber\\
\end{align}
and projecting $H$ [Eq.~(\ref{ham4})] onto the above states, we obtain Eq.~(\ref{Heff}) which we reproduce here,
\begin{equation}
    H_\textrm{eff}=-\tau^z\cos\theta-\tau^x\eta^x\sin\theta\cos\phi+\tau^x\eta^y\sin\theta\sin\phi\,.
\end{equation}

For the ten qubit case, we use the basis states $|\psi_{m_1m_2m_3m_4m_5m_6m_7m_8m_9m_{10}}\rangle$, which are eigenstates of $n$, $n'$, $h$, $h'$, $h^a$, $W_1$, $W_2$, $W_4$, $W_5$, and $W_6$ with eigenvalues $m_1$, $m_2$, $m_3$, $m_4$, $m_5$, $m_6$, $m_7$, $m_8$, $m_9$, and $m_{10}$, respectively. When changing the right clock arm, we project onto the basis states $|\psi_{-1-1-1-1-1-1-1-1-1-1}\rangle$, $|\psi_{-11-1-1-1-1-1-1-1-1}\rangle$, $|\psi_{-1-1-11-1-1-1-1-1-1}\rangle$, and $|\psi_{-11-11-1-1-1-1-1-1}\rangle$. Eq.~(\ref{Heff'}), i.e. $H_\textrm{eff}'$, is obtained by projecting $H_{10}$ onto these states. Then, $\tau^{\alpha\prime}$ and $\eta^{\alpha\prime}$ are defined analogously to $\tau^\alpha$ and $\eta^\alpha$, respectively, upon substituting $|\psi_{mn-1-1}\rangle\rightarrow|\psi_{-1m-1n-1-1-1-1-1-1}\rangle$. Because both $n$ and $n'$ change when changing the middle clock arm, we must project onto eight basis states: $|\psi_{mn-1-1p-1-1-1-1-1}\rangle$ with $m,n,p$ taking values $\pm1$. Projecting $H_{10}$ onto these eight basis states and defining $\chi^\alpha$ analogous to $\tau^{\alpha\prime}$ with the substitution $|\psi_{mn-1p-1-1-1-1-1-1}\rightarrow|\psi_{mn-1-1p-1-1-1-1-1}\rangle$, we obtain $H_\textrm{eff}^a$. 

The gauge fields can be readily obtained from the effective actions using the definitions derived in Ref.~\onlinecite{zeePRA88}. For instance, in order to derive $\mathcal A_\theta$ and $\mathcal A_\phi$, we find two unitary transformations,
\begin{align}
U_y(\theta)&=\left(\begin{array}{cccc}
0&i\cos(\theta/2)&\sin(\theta/2)&0\\
-\cos(\theta/2)&0&0&-i\sin(\theta/2)\\
i\sin(\theta/2)&0&0&\cos(\theta/2)\\
0&-i\sin(\theta/2)&\cos(\theta/2)&0
\end{array}\right)\,,\nonumber
\\
U_z(\phi)&=\exp(i\phi\eta^z/2)\,,
\end{align}
such that $H_\textrm{eff}=-U_z(\phi)^\dagger U_y(\theta)^\dagger\tau_z U_y(\theta)U_z(\phi)$ and $U_y(\theta)$ is in the basis $(|\psi_{-1-1-1-1}\rangle,|\psi_{1-1-1-1}\rangle,|\psi_{-11-1-1}\rangle,|\psi_{11-1-1}\rangle)$. The gauge fields can be calculated from these unitary operations according to $\mathcal A_\theta=U_z(\phi)U_y(\theta)\partial_\theta[U_z(\phi)U_y(\theta)]^\dagger$ and $\mathcal A_\phi=U_z(\phi)U_y(\theta)\partial_\phi[U_z(\phi)U_y(\theta)]^\dagger$. Using an analogous methodology, albeit with different unitary transformations, one can likewise calculate $\mathcal A_{\theta'}$, $\mathcal A_{\phi'}$, $\mathcal A_{\alpha}$, and $\mathcal A_\beta$.

\section{10 MZM Simulation Initialisation Cricuits}
\label{app:10_init_Circuits}

Here, all the circuits to initialise the logical states of the 2-topological qubit, 10-MZM system are given for $q_0$ (Fig.~\ref{fig:Logical_Init_Y2_Q0}) and $q_1$ (Fig.~\ref{fig:Logical_Init_Y2_Q1}).

\begin{figure}
    [t]
    \[\Qcircuit @C=1em @R=1em @!R {
    \lstick{(a)\text{        }\ket{0}}& \gate{H}& \ctrl{1} & \qw &\qw &\qw & \qw &\qw &\qw  \\
    \lstick{\ket{0}}& \gate{X}& \targ & \qw &\qw &\qw & \qw &\qw &\qw \\
    \lstick{\ket{0}}& \gate{X}& \gate{S^\dagger}& \gate{Y} &\targ& \qw & \qw&\qw &\qw && \raisebox{-3em}{$\ket{0}^{q_0}_L$} \\
     \lstick{\ket{0}}& \gate{H} & \qw & \ctrl{-1} &\qw & \targ & \gate{Z}&\qw &\qw  \\
     \lstick{\ket{0}}&  \gate{H} & \qw & \qw &\ctrl{-2} & \ctrl{-1} & \ctrl{1} & \gate{Y} &\qw  \\
     \lstick{\ket{0}}&  \qw & \qw &\qw &\qw &\qw & \targ &\qw &\qw &
    } \]

    \[\Qcircuit @C=1em @R=1em @!R {
    \lstick{(b)\text{        }\ket{0}}& \gate{H}&\gate{Z}& \ctrl{1} & \qw &\qw &\qw & \qw   \\
    \lstick{\ket{0}}& \gate{X}& \qw & \targ & \qw &\qw &\qw & \qw   \\
    \lstick{\ket{0}}& \qw &  \gate{Y} &\targ& \qw & \qw&\qw &\qw && \raisebox{-3em}{$\ket{1}^{q_0}_L$} \\
     \lstick{\ket{0}}& \gate{H}  & \ctrl{-1} &\qw & \targ & \gate{Z}&\qw &\qw  \\
     \lstick{\ket{0}}&  \gate{H} & \qw &\ctrl{-2} & \ctrl{-1} & \ctrl{1} & \gate{Y} &\qw  \\
     \lstick{\ket{0}}&  \qw  &\qw &\qw &\qw & \targ &\qw &\qw &
    } \]

    \[\Qcircuit @C=1em @R=1em @!R {
    \lstick{(c)\text{        }\ket{0}}& \gate{H}&\gate{Z}& \ctrl{1} & \qw &\qw &\ctrl{2} & \ctrl{3} & \qw   \\
    \lstick{\ket{0}}& \gate{X}& \qw & \targ & \qw &\qw &\qw & \qw & \qw  \\
    \lstick{\ket{0}}& \gate{H}& \gate{S} & \gate{Y} &\targ& \gate{Z} & \ctrl{-2}&\qw &\qw && \raisebox{-3em}{$\ket{+}^{q_0}_L$} \\
     \lstick{\ket{0}}& \gate{H} & \qw & \ctrl{-1} &\qw & \targ & \qw &\ctrl{-3} &\qw  \\
     \lstick{\ket{0}}&  \gate{H} & \qw & \qw &\ctrl{-2} & \ctrl{-1} & \ctrl{1} & \gate{Y} &\qw  \\
     \lstick{\ket{0}}&  \qw & \qw &\qw &\qw &\qw & \targ &\qw &\qw &
    } \]

    \[\Qcircuit @C=1em @R=1em @!R {
    \lstick{(d)\text{        }\ket{0}}& \gate{H}&\gate{Z}& \ctrl{1}  &\qw &\ctrl{2} & \ctrl{3} & \qw   \\
    \lstick{\ket{0}}& \gate{X}& \qw & \targ  &\qw &\qw & \qw & \qw  \\
    \lstick{\ket{0}}& \gate{H} & \gate{Y} &\targ& \gate{Z} & \ctrl{-2}&\qw &\qw && \raisebox{-3em}{$\ket{i^+}^{q_0}_L$} \\
     \lstick{\ket{0}}& \gate{H}  & \ctrl{-1} &\qw & \targ & \qw &\ctrl{-3} &\qw  \\
     \lstick{\ket{0}}&  \gate{H} & \qw &\ctrl{-2} & \ctrl{-1} & \ctrl{1} & \gate{R_y(-\pi)} &\qw  \\
     \lstick{\ket{0}}&  \qw  &\qw &\qw &\qw & \targ &\qw &\qw &
    } \]

\caption{Circuits to initialize the required logical states (a) $\ket{0}^{q_0}_L$, (b) $\ket{1}^{q_0}_L$, (c) $\ket{+}^{q_0}_L$ and (d) $\ket{i^+}^{q_0}_L$, for process tomography for the 10-MZM, 2 simulated topological qubit experiments.} 
\label{fig:Logical_Init_Y2_Q0}
\end{figure}
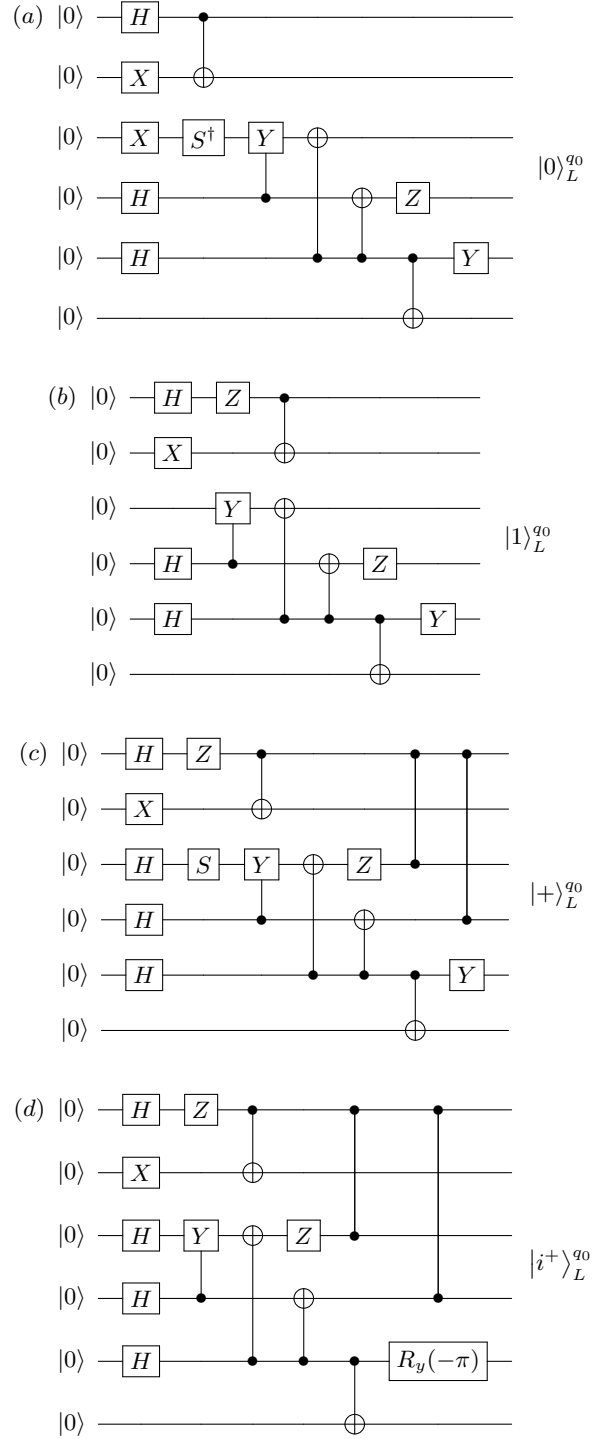

\begin{figure}
    [t]
    \[\Qcircuit @C=1em @R=1em @!R {
    \lstick{(a)\text{        }\ket{0}}& \gate{\sqrt{X}}& \ctrl{2}& \targ & \ctrl{3} & \ctrl{2} & \qw & \qw  \\
    \lstick{\ket{0}}& \gate{H} & \qw & \ctrl{-1} & \qw & \qw & \qw & \qw && \raisebox{-3em}{$\ket{0}^{q_1}_L$} \\
     \lstick{\ket{0}}& \gate{X} & \targ & \qw & \qw & \targ& \qw & \qw   \\
     \lstick{\ket{0}}&  \gate{X} & \qw & \qw & \targ & \qw & \qw & \qw 
    } \]

    \[\Qcircuit @C=1em @R=1em @!R {
    \lstick{(b)\text{        }\ket{0}}& \gate{\sqrt{X}}& \ctrl{2}& \targ & \ctrl{3} & \qw & \ctrl{2} & \qw & \qw  \\
    \lstick{\ket{0}}& \gate{H} & \qw & \ctrl{-1} & \qw & \gate{Z} & \qw & \qw & \qw && \raisebox{-3em}{$\ket{1}^{q_1}_L$} \\
     \lstick{\ket{0}}& \gate{X} & \targ & \qw & \qw &\qw & \targ& \qw & \qw   \\
     \lstick{\ket{0}}&  \qw & \qw & \qw & \targ & \qw & \qw & \qw & \qw 
    } \]

    \[\Qcircuit @C=1em @R=1em @!R {
    \lstick{(c)\text{        }\ket{0}}& \gate{\sqrt{X}}& \ctrl{2}& \gate{Y} &\qw& \ctrl{3} & \ctrl{2} & \qw & \qw & \qw & \qw  \\
    \lstick{\ket{0}}& \gate{H} & \qw & \ctrl{-1} &\gate{S} & \qw & \qw & \qw & \ctrl{2} & \qw & \qw  && \raisebox{-3em}{$\ket{+}^{q_1}_L$} \\
     \lstick{\ket{0}}& \gate{X} & \targ & \qw &\qw & \qw & \targ& \qw & \qw & \gate{Z} & \qw   \\
     \lstick{\ket{0}}&  \gate{H} & \qw & \qw &\qw & \targ & \qw & \qw & \ctrl{-2} & \qw & \qw 
    } \]
    \[\Qcircuit @C=1em @R=1em @!R {
    \lstick{(d)\text{        }\ket{0}}& \gate{\sqrt{X}}& \ctrl{2}& \gate{Y} &\qw& \ctrl{3} & \ctrl{2} & \qw & \qw & \qw & \qw  \\
    \lstick{\ket{0}}& \gate{H} & \qw & \ctrl{-1} &\gate{S} & \qw & \qw & \qw & \ctrl{2} & \qw & \qw  && \raisebox{-3em}{$\ket{i^+}^{q_1}_L$} \\
     \lstick{\ket{0}}& \gate{X} & \targ & \qw &\qw & \qw & \targ& \qw & \qw & \gate{Z} & \qw   \\
     \lstick{\ket{0}}&  \gate{H} & \gate{S^\dagger} & \qw &\qw & \targ & \qw & \qw & \ctrl{-2} & \qw & \qw 
    } \]
\caption{Circuits to initialize the required logical states (a) $\ket{0}^{q_1}_L$, (b) $\ket{1}^{q_1}_L$, (c) $\ket{+}^{q_1}_L$ and (d) $\ket{i^+}^{q_1}_L$, for process tomography for the 10-MZM, 2 simulated topological qubit experiments.} 
\label{fig:Logical_Init_Y2_Q1}
\end{figure}

\section{Simulation Data}
\label{app:Sim_Data}

Here, the results of the classical simulations and \textit{ibm\_brisbane} demonstrations of each circuit tested in our single and two-topological qubit quantum simulations are given as state fidelities (Tab.~\ref{tab:TabFull_Brisbane_StateTomo}-~\ref{tab:TabFull_Brisbane_10qubit_MinusPi2}). Additionally, example Hinton plots of select single (Fig.~\ref{fig:Hinton_S}) and entangling (Fig.~\ref{fig:Hinton_RX}) simulated topological qubit operations are given. 

\begin{figure}
    \centering
    \includegraphics[width=0.45\linewidth]{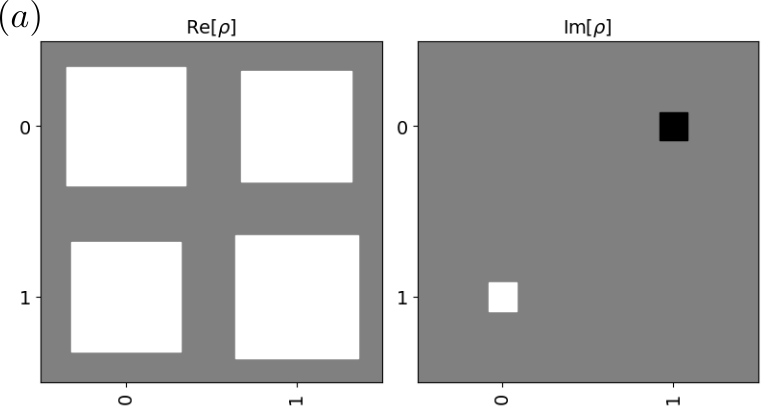} \hspace{2em}
    \includegraphics[width=0.45\linewidth]{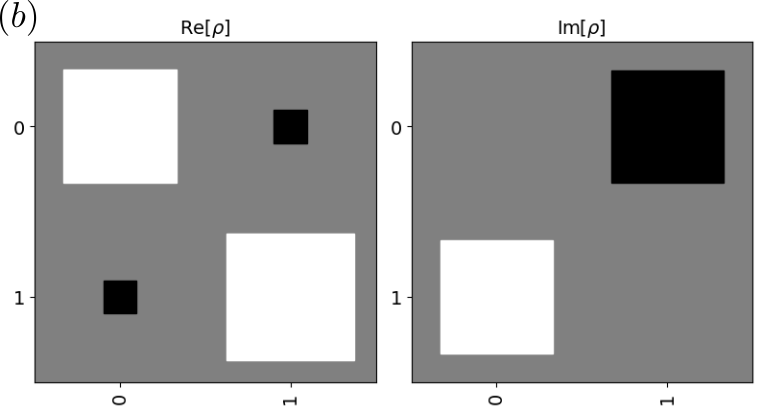}
    \caption{Example Hinton plots of the real and imaginary components of the density matrices constructed from the (a)$\ket{+}_L$ and (b)$S\ket{+}_L$ circuits as demonstrated on the \textit{ibm\_brisbane} QPU in $2^{13}$ shot runs. White (black) blocks indicate positive (negative) values of the relevant matrix elements whilst the block size indicates their magnitude.}
    \label{fig:Hinton_S}
\end{figure}

\begin{figure}
    \centering
    \includegraphics[width=0.75\linewidth]{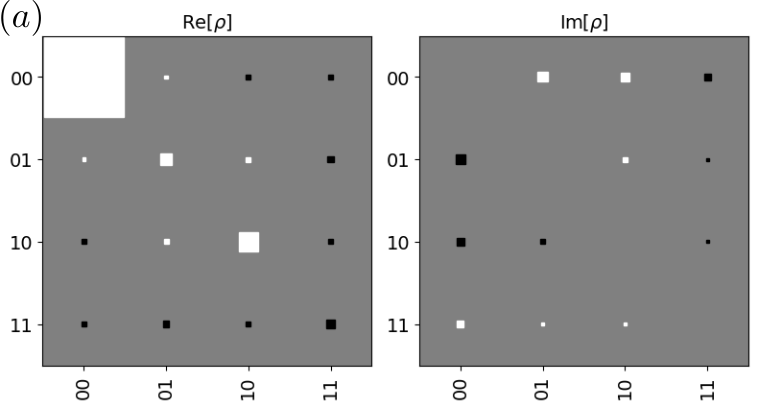}
    \includegraphics[width=0.75\linewidth]{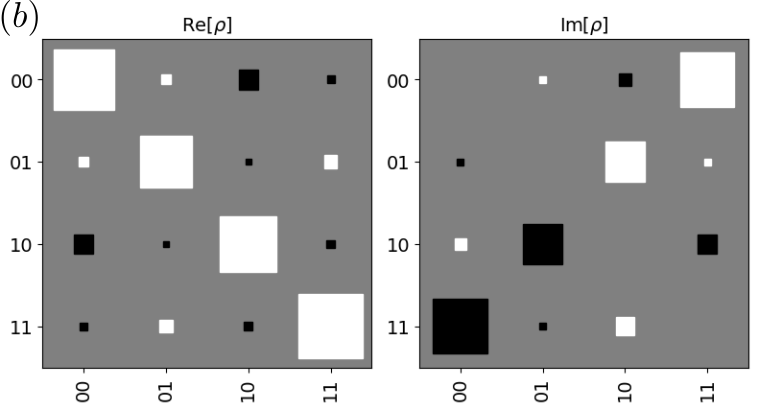}
    \caption{Example Hinton plots of the real and imaginary components of the density matrices constructed from the (a)$\ket{0}_L^{q_0}\ket{0}_L^{q_1}$ and (b)$R_{xx}\left(\frac{\pi}{2}\right)\ket{0}_L^{q_0}\ket{0}_L^{q_1}$ circuits as demonstrated on the \textit{ibm\_brisbane} QPU in $2^{13}$ shot runs. White (black) blocks indicate positive (negative) values of the relevant matrix elements whilst the block size indicates their magnitude.}
    \label{fig:Hinton_RX}
\end{figure}

\begin{table}
    \centering
    \begin{tabular}{|l||c|c|c|c|}
        \hline
         Operation & Simulation  &  Experiment &  Depth &  $|\tilde\Delta|$ \\
         \hline \hline
        $\ket{0}_L$ & $97.07\pm0.26\%$  & $99.22\pm0.56\%$ & 35 & - \\
        $\ket{1}_L$ & $91.38\pm0.44\%$  & $96.95\pm0.59\%$ & 16 & - \\
        $\ket{+}_L$ & $95.02\pm0.34\%$  & $91.54\pm0.58\%$ & 28 & - \\
        $\ket{i^+}_L$ & $93.16\pm0.39\%$  & $82.85\pm0.49\%$ & 34 & - \\
        \hline 
        $S\ket{0}_L$ & $87.70\pm0.51 \%$ & $77.82\pm0.52 \%$ & 139  & 6.3   \\
        $S\ket{1}_L$ & $86.18 \pm0.54 \%$ & $73.47\pm0.49 \%$ &  128 & 6.3 \\
        $S\ket{+}_L$ & $93.80\pm0.38 \%$ & $82.57\pm0.48 \%$ & 132  & 6.3 \\
        $S\ket{i^+}_L$ & $86.08\pm0.54 \%$ & $80.42\pm4.31 \%$ &  147 & 6.3 \\
        \hline 
        $T\ket{0}_L$ & $79.48\pm0.45\%$ & $61.55\pm0.65\%$ & 308 & 4.2 \\
        $T\ket{1}_L$ & $77.85\pm0.46\%$ & $65.74\pm0.57\%$ & 290 & 4.2 \\
        $T\ket{+}_L$ & $84.23\pm0.67\%$ & $65.25\pm0.86\%$ & 295 & 4.2 \\
        $T\ket{i^+}_L$ & $60.13\pm0.87\%$ & $60.41\pm3.49\%$ & 313 & 4.2 \\
        \hline 
        $S^{\dagger}\ket{0}_L$ & $90.36\pm0.46\%$ & $74.95\pm0.60\%$ & 199 & 7.0 \\
        $S^{\dagger}\ket{1}_L$ & $87.33\pm0.52\%$ & $66.20\pm0.68\%$ & 173 & 7.0 \\
        $S^{\dagger}\ket{+}_L$ & $91.02\pm0.45\%$ & $86.40\pm0.53\%$ & 173 & 7.0 \\
        $S^{\dagger}\ket{i^+}_L$ & $89.50\pm0.48\%$ & $82.48\pm0.65\%$ & 185 & 7.0 \\
        \hline 
        $T^{\dagger}\ket{0}_L$ & $82.01\pm0.42 \%$  & $68.83\pm3.44 \%$ & 305 & 4.0 \\
        $T^{\dagger}\ket{1}_L$ & $83.69 \pm0.40 \%$ & $83.30\pm3.86 \%$ & 288 & 4.0 \\
        $T^{\dagger}\ket{+}_L$ & $83.62 \pm0.67 \%$ & $64.74\pm1.03 \%$ & 294 & 4.0 \\
        $T^{\dagger}\ket{i^+}_L$ & $80.95 \pm0.68 \%$ & $70.72\pm1.40 \%$ & 403 & 4.0 \\
        \hline
    \end{tabular}
    \caption{Initialization fidelity and state tomography simulation results, experimental results, circuit depth and associated $|\tilde\Delta|$ for the simulated 4-MZM, simulated topological qubit logical basis on the \textit{ibm\_brisbane} quantum processor. Here, each experiment consists of $2^{13}$ shots.}
    \label{tab:TabFull_Brisbane_StateTomo}
\end{table}

\begin{table}
    \centering
    \begin{tabular}{|l||c|c|c|}
        \hline
         Operation & Simulation  &  Experiment &  Depth  \\
         \hline \hline
        $\ket{0}_L^{q_0}\ket{0}_L^{q_1}$ & $90.53\pm0.56\%$  & $92.28\pm1.11\%$ & 33  \\
        $\ket{0}_L^{q_0}\ket{+}_L^{q_1}$ & $82.92\pm0.72\%$  & $94.13\pm1.13\%$ & 33  \\
        $\ket{0}_L^{q_0}\ket{i^+}_L^{q_1}$ & $90.89\pm0.55\%$  & $94.14\pm1.00\%$ & 33  \\
        $\ket{0}_L^{q_0}\ket{1}_L^{q_1}$ & $89.07\pm0.60\%$  & $94.84\pm0.91\%$ & 30  \\
        \hline
        $\ket{+}_L^{q_0}\ket{0}_L^{q_1}$ & $89.65\pm0.59\%$  & $81.44\pm1.35\%$ & 37  \\
        $\ket{+}_L^{q_0}\ket{+}_L^{q_1}$ & $86.08\pm0.66\%$  & $88.16\pm1.45\%$ & 37 \\
        $\ket{+}_L^{q_0}\ket{i^+}_L^{q_1}$ & $88.65\pm0.60\%$  & $88.21\pm1.23\%$ & 41 \\
        $\ket{+}_L^{q_0}\ket{1}_L^{q_1}$ & $89.79\pm0.58\%$  & $81.15\pm1.39\%$ & 39  \\
        \hline
        $\ket{i^+}_L^{q_0}\ket{0}_L^{q_1}$ & $86.57\pm0.66\%$  & $80.77\pm1.29\%$ & 32  \\
        $\ket{i^+}_L^{q_0}\ket{+}_L^{q_1}$ & $85.41\pm0.68\%$  & $89.52\pm1.23\%$ & 35 \\
        $\ket{i^+}_L^{q_0}\ket{i^+}_L^{q_1}$ & $85.79\pm0.67\%$  & $89.59\pm1.17\%$ & 35  \\
        $\ket{i^+}_L^{q_0}\ket{1}_L^{q_1}$ & $87.63\pm0.63\%$  & $82.24\pm1.38\%$ & 35  \\
        \hline
        $\ket{1}_L^{q_0}\ket{0}_L^{q_1}$ & $88.51\pm0.61\%$  & $91.89\pm0.97\%$ & 30  \\
        $\ket{1}_L^{q_0}\ket{+}_L^{q_1}$ & $86.55\pm0.64\%$  & $94.70\pm1.12\%$ & 33  \\
        $\ket{1}_L^{q_0}\ket{i^+}_L^{q_1}$ & $87.35\pm0.63\%$  & $94.52\pm0.81\%$ & 36  \\
        $\ket{1}_L^{q_0}\ket{1}_L^{q_1}$ & $90.55\pm0.56\%$  & $91.77\pm0.99\%$ & 26  \\
        \hline
    \end{tabular}
    \caption{Initialization fidelity simulation results, experimental results and circuit depth for the simulated 10-MZM, 2 simulated topological qubit logical basis on the \textit{ibm\_brisbane} quantum processor. Here, each experiment consists of $2^{13}$ shots.}
    \label{tab:TabFull_Brisbane_10qubit_Init}
\end{table}

\begin{table}
    \centering
    \begin{tabular}{|l||c|c|c|c|}
        \hline
         Operation & Simulation  &  Experiment &  Depth \\
         \hline \hline
        $R_{xx}\left(\frac{\pi}{2}\right)\ket{0}_L^{q_0}\ket{0}_L^{q_1}$ & $67.64\pm0.67\%$  & $50.18\pm0.97\%$ & 130   \\
        $R_{xx}\left(\frac{\pi}{2}\right)\ket{0}_L^{q_0}\ket{+}_L^{q_1}$ & $78.68\pm0.57\%$  & $73.28\pm0.85\%$ & 141  \\
        $R_{xx}\left(\frac{\pi}{2}\right)\ket{0}_L^{q_0}\ket{i^+}_L^{q_1}$ & $63.92\pm0.70\%$  & $60.43\pm1.03\%$ & 158  \\
        $R_{xx}\left(\frac{\pi}{2}\right)\ket{0}_L^{q_0}\ket{1}_L^{q_1}$ & $64.63\pm0.70\%$  & $59.73\pm0.80\%$ & 153  \\
        \hline
        $R_{xx}\left(\frac{\pi}{2}\right)\ket{+}_L^{q_0}\ket{0}_L^{q_1}$ & $65.23\pm0.70\%$  & $64.34\pm0.81\%$ & 146  \\
        $R_{xx}\left(\frac{\pi}{2}\right)\ket{+}_L^{q_0}\ket{+}_L^{q_1}$ & $63.04\pm0.67\%$  & $63.84\pm0.84\%$ & 147 \\
        $R_{xx}\left(\frac{\pi}{2}\right)\ket{+}_L^{q_0}\ket{i^+}_L^{q_1}$ & $66.31\pm0.69\%$  & $63.05\pm0.86\%$ & 144 \\
        $R_{xx}\left(\frac{\pi}{2}\right)\ket{+}_L^{q_0}\ket{1}_L^{q_1}$ & $67.96\pm0.68\%$  & $58.77\pm0.89\%$ & 148  \\
        \hline
        $R_{xx}\left(\frac{\pi}{2}\right)\ket{i^+}_L^{q_0}\ket{0}_L^{q_1}$ & $66.75\pm0.69\%$  & $57.68\pm0.92\%$ & 174  \\
        $R_{xx}\left(\frac{\pi}{2}\right)\ket{i^+}_L^{q_0}\ket{+}_L^{q_1}$ & $70.66\pm0.64\%$  & $57.14\pm0.88\%$ & 176 \\
        $R_{xx}\left(\frac{\pi}{2}\right)\ket{i^+}_L^{q_0}\ket{i^+}_L^{q_1}$ & $66.97\pm0.69\%$  & $58.67\pm0.94\%$ & 141  \\
        $R_{xx}\left(\frac{\pi}{2}\right)\ket{i^+}_L^{q_0}\ket{1}_L^{q_1}$ & $62.29\pm0.72\%$  & $55.10\pm1.01\%$ & 178  \\
        \hline
        $R_{xx}\left(\frac{\pi}{2}\right)\ket{1}_L^{q_0}\ket{0}_L^{q_1}$ & $65.31\pm0.69\%$  & $61.98\pm0.89\%$ & 150  \\
        $R_{xx}\left(\frac{\pi}{2}\right)\ket{1}_L^{q_0}\ket{+}_L^{q_1}$ & $76.80\pm0.57\%$  & $65.03\pm0.87\%$ & 123  \\
        $R_{xx}\left(\frac{\pi}{2}\right)\ket{1}_L^{q_0}\ket{i^+}_L^{q_1}$ & $65.61\pm0.69\%$  & $58.61\pm0.95\%$ & 135  \\
        $R_{xx}\left(\frac{\pi}{2}\right)\ket{1}_L^{q_0}\ket{1}_L^{q_1}$ & $62.86\pm0.71\%$  & $62.11\pm0.86\%$ & 148  \\
        \hline
    \end{tabular}
    \caption{State tomography fidelity simulation results, experimental results and circuit depth for the simulated $R_{xx}\left(\frac{\pi}{2}\right)$ braiding gate on the \textit{ibm\_brisbane} quantum processor. Here, each experiment consists of $2^{13}$ shots and $|\tilde\Delta|=6.3$.}
    \label{tab:TabFull_Brisbane_10qubit_PlusPi2}
\end{table}

\begin{table}
    \centering
    \begin{tabular}{|l||c|c|c|c|}
        \hline
         Operation & Simulation  &  Experiment &  Depth \\
         \hline \hline
        $R_{xx}\left(-\frac{\pi}{2}\right)\ket{0}_L^{q_0}\ket{0}_L^{q_1}$ & $72.34\pm0.64\%$  & $59.58\pm0.89\%$ & 160   \\
        $R_{xx}\left(-\frac{\pi}{2}\right)\ket{0}_L^{q_0}\ket{+}_L^{q_1}$ & $74.87\pm0.60\%$  & $62.34\pm0.83\%$ & 165  \\
        $R_{xx}\left(-\frac{\pi}{2}\right)\ket{0}_L^{q_0}\ket{i^+}_L^{q_1}$ & $74.29\pm0.62\%$  & $64.40\pm0.96\%$ & 159  \\
        $R_{xx}\left(-\frac{\pi}{2}\right)\ket{0}_L^{q_0}\ket{1}_L^{q_1}$ & $75.55\pm0.61\%$  & $61.97\pm0.76\%$ & 169  \\
        \hline
        $R_{xx}\left(-\frac{\pi}{2}\right)\ket{+}_L^{q_0}\ket{0}_L^{q_1}$ & $77.98\pm0.59\%$  & $67.05\pm0.81\%$ & 156  \\
        $R_{xx}\left(-\frac{\pi}{2}\right)\ket{+}_L^{q_0}\ket{+}_L^{q_1}$ & $78.40\pm0.57\%$  & $72.77\pm0.70\%$ & 168 \\
        $R_{xx}\left(-\frac{\pi}{2}\right)\ket{+}_L^{q_0}\ket{i^+}_L^{q_1}$ & $74.84\pm0.62\%$  & $62.50\pm0.96\%$ & 182 \\
        $R_{xx}\left(-\frac{\pi}{2}\right)\ket{+}_L^{q_0}\ket{1}_L^{q_1}$ & $71.32\pm0.65\%$  & $66.81\pm0.85\%$ & 173  \\
        \hline
        $R_{xx}\left(-\frac{\pi}{2}\right)\ket{i^+}_L^{q_0}\ket{0}_L^{q_1}$ & $68.89\pm0.67\%$  & $58.22\pm0.91\%$ & 175  \\
        $R_{xx}\left(-\frac{\pi}{2}\right)\ket{i^+}_L^{q_0}\ket{+}_L^{q_1}$ & $75.32\pm0.61\%$  & $64.32\pm0.87\%$ & 170 \\
        $R_{xx}\left(-\frac{\pi}{2}\right)\ket{i^+}_L^{q_0}\ket{i^+}_L^{q_1}$ & $74.23\pm0.62\%$  & $62.40\pm0.89\%$ & 170  \\
        $R_{xx}\left(-\frac{\pi}{2}\right)\ket{i^+}_L^{q_0}\ket{1}_L^{q_1}$ & $75.38\pm0.61\%$  & $63.63\pm0.88\%$ & 170  \\
        \hline
        $R_{xx}\left(-\frac{\pi}{2}\right)\ket{1}_L^{q_0}\ket{0}_L^{q_1}$ & $74.61\pm0.62\%$  & $56.23\pm0.94\%$ & 161  \\
        $R_{xx}\left(-\frac{\pi}{2}\right)\ket{1}_L^{q_0}\ket{+}_L^{q_1}$ & $79.86\pm0.56\%$  & $63.49\pm0.84\%$ & 164  \\
        $R_{xx}\left(-\frac{\pi}{2}\right)\ket{1}_L^{q_0}\ket{i^+}_L^{q_1}$ & $61.44\pm0.72\%$  & $49.47\pm0.94\%$ & 166  \\
        $R_{xx}\left(-\frac{\pi}{2}\right)\ket{1}_L^{q_0}\ket{1}_L^{q_1}$ & $62.52\pm0.72\%$  & $45.82\pm1.01\%$ & 159  \\
        \hline
    \end{tabular}
    \caption{State tomography fidelity simulation results, experimental results and circuit depth for the simulated $R_{xx}\left(-\frac{\pi}{2}\right)$ braiding gate on the \textit{ibm\_brisbane} quantum processor. Here, each experiment consists of $2^{13}$ shots and $|\tilde\Delta|=7.2$.}
    \label{tab:TabFull_Brisbane_10qubit_MinusPi2}
\end{table}

\end{widetext}
\end{document}